\documentclass[12pt]{article}

\usepackage[a4paper,text={16.3cm,22cm}]{geometry}
\usepackage{amsmath,amsfonts,braket,slashed,amssymb,bm,bbm,xcolor,graphicx}
\usepackage[small,labelfont=bf]{caption}
\usepackage{cite}
\usepackage[
    bookmarksnumbered=true,
    urlcolor=blue,
    linkbordercolor=red,
    citebordercolor=green,
    bookmarksopen=true
    ]{hyperref}

\allowdisplaybreaks
\numberwithin{equation}{section}

\newcommand{\GGammac}{{\rm{I}}\hspace{-0.8mm}\Gamma^c}
\newcommand{\spac}{{\hspace{0.3mm}}}

\title{
Renormalization-Group Improved Resummation
\texorpdfstring{\\[1ex]}{}
of Super-Leading Logarithms
}

\begin{document}

\begin{titlepage}

\begin{flushright}
{\small
MITP-24-012\\
May 8, 2024
}
\end{flushright}

\makeatletter
\vskip0.8cm
\pdfbookmark[0]{\@title}{title}
\begin{center}
{\Large \bf\boldmath \@title}
\end{center}
\makeatother

\vspace{0.5cm}
\begin{center}
\textsc{Philipp Böer,$^a$ Patrick Hager,$^a$ Matthias Neubert,$^{a,b}$ \\ 
Michel Stillger$^a$ and Xiaofeng Xu$^a$} \\[6mm]
    
\textsl{${}^a$PRISMA$^+$ Cluster of Excellence \& Mainz Institute for Theoretical Physics\\
Johannes Gutenberg University, Staudingerweg 9, 55128 Mainz, Germany\\[0.3cm]
${}^b$Department of Physics \& LEPP, Cornell University, Ithaca, NY 14853, U.S.A.}
\end{center}

\vspace{0.6cm}
\pdfbookmark[1]{Abstract}{abstract}
\begin{abstract}
A new strategy is presented for systematically treating super-leading logarithmic contributions including higher-order Glauber exchanges for non-global LHC observables in renormalization-group (RG) improved perturbation theory. This represents an important improvement over previous approaches, as it allows for the consistent inclusion of the scale dependence of the strong coupling, thereby providing more reliable estimates of the scale uncertainties in theoretical predictions. The key idea is to rearrange the relevant RG evolution operator in such a way that all double-logarithmic corrections are exponentiated from the outset. This forms the starting point for the first resummation of super-leading logarithms at leading order in RG-improved perturbation theory for arbitrary $2\to M$ scattering processes. Moreover, the asymptotic scaling of subleading logarithmic corrections from higher-order Glauber exchanges is determined, demonstrating their parametric suppression.
\end{abstract}

\vfill\noindent\rule{0.4\columnwidth}{0.4pt}\\
\hspace*{2ex} {\small \textit{E-mail:} \href{mailto:pboeer@uni-mainz.de}{pboeer@uni-mainz.de}, \href{mailto:pahager@uni-mainz.de}{pahager@uni-mainz.de}, \href{mailto:matthias.neubert@uni-mainz.de}{matthias.neubert@uni-mainz.de}, \\
\hspace*{2ex} \phantom{E-mail: } \href{mailto:m.stillger@uni-mainz.de}{m.stillger@uni-mainz.de}, \href{mailto:xiaxu@uni-mainz.de}{xiaxu@uni-mainz.de}}

\end{titlepage}

{\hypersetup{hidelinks}
\pdfbookmark[1]{Contents}{ToC}
\setcounter{tocdepth}{1}
\tableofcontents}
\vspace{6mm}

\section{Introduction}

Precision studies of jet observables at high-energy hadron colliders, such as the CERN Large Hadron Collider (LHC), play an important role in the pursuit to test the Standard Model of elementary-particle physics and look for hints of physics that lies beyond it. The reason is that jet processes closely resemble the underlying hard-scattering dynamics, thereby providing an opportunity to study the fundamental interactions at shortest distances. On the other hand, jet rates are among the most difficult observables to calculate theoretically. The distinction of the highly collimated energetic particles constituting the jets from the soft out-of-jet radiation by imposing veto criteria in certain regions of phase space makes these observables ``non-global''~\cite{Sterman:1977wj}. For gap-between-jets cross sections, for instance, the energy emitted into a region away from the jets and the beam remnants is restricted by introducing a jet-veto scale $Q_0$ much smaller than the hard scale $Q\sim\sqrt{\hat{s}}$ set by the partonic center-of-mass energy. Such vetoes introduce additional scales in the problem, which in turn give rise to large logarithmic corrections, whose structure can be very complicated. So-called ``non-global logarithms'' (NGLs) arise from soft-gluon radiation off secondary emissions inside the jets~\cite{Dasgupta:2001sh}. An all-order resummation of these logarithms is highly non-trivial due to their intricate pattern and the complexity of the color algebra involved. At $e^+ e^-$ colliders, and working in the large-$N_c$ approximation, the resummation of the leading NGLs can be accomplished by solving a non-linear integro-differential equation derived in~\cite{Banfi:2002hw}. The resummation of NGLs at finite $N_c$ has been studied in~\cite{Weigert:2003mm,Hatta:2013iba,Hatta:2020wre,DeAngelis:2020rvq}, and the first resummations at next-to-leading logarithmic accuracy have been achieved in~\cite{Banfi:2021owj,Banfi:2021xzn,Becher:2021urs,Becher:2023vrh,FerrarioRavasio:2023kyg}.

In hadronic scattering, Glauber-gluon exchanges between the initial-state partons can lead to a violation of collinear factorization~\cite{Catani:2011st,Forshaw:2012bi,Schwartz:2017nmr}. This implies a breakdown of color coherence in higher orders of perturbation theory, resulting in the appearance of so-called ``super-leading logarithms'' (SLLs), an infinite series of double-logarithmic corrections in the scale ratio $Q/Q_0\gg 1$~\cite{Forshaw:2006fk,Forshaw:2008cq,Keates:2009dn}. Their all-order resummation has been accomplished using an effective field-theory framework, which allows for a systematic theoretical description of non-global observables at hadron colliders~\cite{Becher:2021zkk,Becher:2023mtx}. It is briefly reviewed in Section~\ref{sec:review}. Working with a fixed coupling $\alpha_s(\bar\mu)$, with $Q_0<\bar\mu<Q$, the corresponding contributions to a partonic $2\to M$ jet cross section, expressed in terms of the Born cross section $\hat\sigma_{2\to M}$, can be written in the schematic form
\begin{equation}\label{eq:SLLgeneric}
   \hat\sigma_{2\to M}^{\rm SLL}(Q_0) 
   = \hat\sigma_{2\to M}\,\frac{\alpha_s(\bar\mu)\spac L}{\pi\spac N_c}\,w_\pi 
    \sum_{n=0}^\infty\,c_{1,n}\,w^{n+1} \,,
\end{equation}
where we have introduced the parameters $w=\frac{N_c\spac\alpha_s(\bar\mu)}{\pi}\,L^2$ and $w_\pi=\frac{N_c\spac\alpha_s(\bar\mu)}{\pi}\,\pi^2$, with the large logarithm $L=\ln(Q/Q_0)\gg 1$. The factor $w_\pi$ contains two Glauber phases, which originate from the imaginary part of $\ln[(-Q^2-i0)/Q_0^2]=2L-i\pi$. In general, the coefficients $c_{1,n}$ depend on the kinematics of the Born process, but we omit this dependence for simplicity. The prefactor in~\eqref{eq:SLLgeneric} shows that the SLLs are subleading in the large-$N_c$ expansion. The all-order resummation of this double-logarithmic series for arbitrary \mbox{$2\to M$} partonic scattering processes has been accomplished in~\cite{Becher:2023mtx}, where the result was expressed in terms of Kamp\'e de F\'eriet functions of the variable $w$. It was shown that in the asymptotic limit $w\gg 1$ the cross section scales like
\begin{equation}
   \hat\sigma_{2\to M}^{\rm SLL}(Q_0) 
   \sim \hat\sigma_{2\to M}\,\frac{\alpha_s(\bar\mu)\spac L}{\pi\spac N_c}\,w_\pi \ln w \,,
\end{equation}
implying an asymptotic behavior vastly different from standard Sudakov form factors. 

For typical scale choices relevant for LHC phenomenology (e.g.\ $Q\sim 0.5$\,--\spac 2\,TeV and $Q_0\sim 10$\,--\,50\,GeV), one finds that the variables $w$ and $w_\pi$ are both of $\mathcal{O}(1)$. The contribution of the series~\eqref{eq:SLLgeneric} is then expected to be as large as a one-loop correction to the Born cross section. Numerical estimates of the SLL contributions to all $2\to 2$ partonic QCD scattering processes confirm this claim~\cite{Becher:2023mtx}. On the other hand, with $w_\pi\sim\mathcal{O}(1)$, it is natural to ask what happens when higher-order insertions of Glauber phases are taken into account, thus resumming terms enhanced by the large numerical factor $\pi^2$. This leads to the alternating ``Glauber series''~\cite{Becher:2023mtx} 
\begin{equation}\label{eq:Glauberseries_generic}
   \hat\sigma_{2\to M}^{\rm SLL+G}(Q_0) 
   = \hat\sigma_{2\to M}\,\frac{\alpha_s(\bar\mu)\spac L}{\pi\spac N_c}
    \sum_{\ell=1}^\infty \sum_{n=0}^\infty\,c_{\ell,n}\,w_\pi^\ell\,w^{n+\ell} \,,
\end{equation}
with $c_{\ell,n}\propto (-1)^{n+\ell}$, and $\ell$ denoting the number of pairs of Glauber phases. The Glauber series resums all double-logarithmic and $\pi^2$-enhanced corrections to the cross section. The calculation of the expansion coefficients $c_{\ell,n}$ to all orders of perturbation theory, and for generic $2\to M$ jet processes, has been achieved in~\cite{Boer:2023jsy,Boer:2023ljq}. However, no closed-form expressions for the resummed Glauber series have been obtained so far. Phenomenologically, it turns out that higher-order Glauber term (with $\ell\ge 2$) give rise to small corrections for most $2\to 2$ partonic scattering processes. Nonetheless, for some processes involving gluons in the initial state, the four-Glauber contribution $\propto w_\pi^2$ can have a sizable effect~\cite{Boer:2023ljq}. 

The techniques developed so far~\cite{Becher:2021zkk,Becher:2023mtx,Boer:2023jsy,Boer:2023ljq} have some limitations, which we aim to overcome in the present work:
\begin{itemize}
\item
While for a fixed coupling $\alpha_s(\bar\mu)$ analytic expressions for the all-order resummation of SLLs have been obtained~\cite{Becher:2023mtx}, no such results are known for higher-order terms in the Glauber series. Up to now, only the coefficients $c_{\ell,n}$ in~\eqref{eq:Glauberseries_generic} have been determined in terms of multiple infinite sums~\cite{Boer:2023jsy,Boer:2023ljq}.
\item
The resummation has not been performed systematically in renormalization-group (RG) improved perturbation theory. This is needed, however, to alleviate the large scale uncertainties of the results, previously estimated by varying $\bar\mu$ in the interval between $Q_0$ and $Q$. In~\cite{Becher:2023mtx} some approximate results have been derived using a one-loop running coupling in the relevant scale integrals for the first few SLLs, but it is well known that a consistent treatment requires the inclusion of two-loop contributions. Moreover, the obtained expressions are so cumbersome that in practice they cannot be extended to very high orders. For higher Glauber-phase insertions, no corresponding results are known.
\item
To consistently combine leading and subleading logarithmic effects, it is crucial to determine the asymptotic scaling for $\alpha_s L^2\to\infty$, corresponding to $w\gg 1$, of the higher-order terms in the Glauber series. Numerical results suggest that these terms are suppressed~\cite{Boer:2023jsy,Boer:2023ljq}. However, the current representation as a perturbative series in $w$ impedes an explicit derivation of a possible parametric suppression.
\end{itemize}

In this paper, we present a new strategy for performing an efficient resummation of SLLs and higher-order terms in the Glauber series. It offers the advantage that all double-logarithmic contributions are resummed to all orders of  perturbation theory and ``exponentiated'' into generalized Sudakov factors. This approach allows for a straightforward inclusion of the running of the strong coupling $\alpha_s(\mu)$, which is the basis for an RG-improved resummation scheme. For quark-initiated scattering processes, we present explicit expressions accomplishing the resummation of SLL terms involving up to four insertions of Glauber phases. After setting up the basic formalism, the new strategy is presented in Section~\ref{sec:Uoperator}. Our approach also makes it possible to study the asymptotic behavior of higher-order terms in the Glauber series in the limit where $\alpha_s\spac L^2\to\infty$. Section~\ref{sec:asymptotics} is devoted to deriving the corresponding scaling relations for quark-initiated scattering processes and provides an explanation of the empirical fact that the Glauber series converges rapidly even in the asymptotic limit. As a by-product, we obtain a quasi-analytical resummation of SLL terms with up to four Glauber insertions in the limit where the running of the strong coupling is neglected. A numerical analysis of our results, including a comparison of the results obtained in RG-improved perturbation theory with those derived using a fixed coupling $\alpha_s(\bar\mu)$, is performed in Section~\ref{sec:numerics}. In Section~\ref{sec:resummation_arbitrary} we generalize our approach to arbitrary scattering processes, focusing however on the case of only two insertions of the Glauber operator. A summary of our main results and concluding remarks are presented in Section~\ref{sec:conclusions}. The treatment of higher-order terms in the Glauber series for processes involving gluons in the initial state is cumbersome and is therefore relegated to Appendix~\ref{app:gluons}. Appendix~\ref{app:Sudakov} contains some technical details for implementing the crossing of heavy-flavor thresholds.

\section{Review of the basic formalism}
\label{sec:review}

The starting point for our analysis is the factorization formula~\cite{Balsiger:2018ezi,Becher:2021zkk,Becher:2023mtx}
\begin{equation}\label{eq:factorization_formula}
   \sigma_{2\to M}(Q_0) 
   = \int\!d\xi_1 \int\!d\xi_2\,\sum_{m=2+M}^\infty 
    \big\langle \bm{\mathcal{H}}_m(\{\underline{n}\},s,\xi_1,\xi_2,\mu) 
    \otimes \bm{\mathcal{W}}_m(\{\underline{n}\},Q_0,\xi_1,\xi_2,\mu) \big\rangle
\end{equation}
for a $2\to M$ jet cross section at a hadron collider. Here, $s$ denotes the squared center-of-mass energy, $\xi_1$ and $\xi_2$ are the longitudinal momentum fractions carried by the colliding partons, and $Q_0$ is the soft scale associated with the veto imposed on radiation between the beam remnants and the final-state jets. The tuple $\{\underline{n}\}=\{n_1,\dots,n_m\}$ collects light-like 4-vectors aligned with the directions of the initial-state ($i=1,2$) and final-state ($i=3,\dots,m$) particles, defined in the laboratory frame. The above formula generalizes an analogous result for $e^+ e^-$ colliders derived in~\cite{Becher:2015hka,Becher:2016mmh}. The hard functions $\bm{\mathcal{H}}_m$ of multiplicity $m$ describe all possible $m$-particle scattering processes $1+2\to 3+\dots+m$, where $i$ represents the $i$-th particle ($i=q,\bar q,g$ for colored partons). To keep the notation compact, we do not indicate the different partonic configurations, but it is understood that one must sum not only over different values of $m$, but over all possible channels. The low-energy matrix elements $\bm{\mathcal{W}}_m$ describe both the soft emissions off the hard partons and the collinear dynamics associated with the initial state. Both $\bm{\mathcal{H}}_m$ and $\bm{\mathcal{W}}_m$ are density matrices in color space~\cite{Catani:1996vz}. The symbol $\otimes$ indicates an integration over the directions $\{n_3,\dots,n_m\}$ of the final-state particles, and the brackets $\langle\dots\rangle$ denote a sum (average) over final-state (initial-state) color and spin indices. Explicit definitions of the various functions entering the factorization formula and of the notation used in our work can be found in~\cite{Becher:2023mtx}.

The jet cross section $\sigma_{2\to M}(Q_0)$ receives large logarithmic corrections in the scale ratio $Q/Q_0$, between the hard scale $Q\sim\sqrt{\hat s}=\sqrt{\xi_1\spac \xi_2\spac s}$ and the low scale $Q_0$.\footnote{In leading-logarithmic approximation, one is insensitive to the precise definition of the jet observable.} 
These corrections can be resummed to all orders of perturbation theory by solving the RG evolution equation of the hard functions, which is of the form~\cite{Becher:2021zkk}
\begin{equation}\label{eq:hardRG}
   \frac{d}{d\ln\mu}\,\bm{\mathcal{H}}_m(\{\underline{n}\},s,\mu)
   = - \sum_{l=2+M}^{m} \bm{\mathcal{H}}_l(\{\underline{n}\},s,\mu)\,\star
    \bm{\Gamma}^H_{lm}(\{\underline{n}\},s,\mu) \,.
\end{equation}
The symbol $\star$ denotes the Mellin convolution over the longitudinal momentum fractions of the initial-state partons. The anomalous dimension in this equation is an operator in color space and in the (infinite) space of parton multiplicities, and this fact makes the evaluation of the solution a very challenging task. Importantly, the multiplicities change under scale evolution. The detailed form of the anomalous dimension has been derived in~\cite{Becher:2023mtx}. The formal solution to the evolution equation can be expressed in terms of the path-ordered exponential 
\begin{equation}\label{eq:path_ordered_exponential}
   \bm{U}(\{\underline{n}\},s,\mu_h,\mu) 
   = \mathbf{P} \exp\biggl[ \int_{\mu}^{\mu_h}\frac{d\mu'}{\mu'}\,\bm{\Gamma}^H(\{\underline{n}\},s,\mu') \biggr]\,.
\end{equation}
Its action on the hard functions is defined by the series expansion 
\begin{align}\label{eq:Uexp}
   \bm{\mathcal{H}}(\{\underline{n}\},s,\mu_s)
   &= \bm{\mathcal{H}}(\{\underline{n}\},s,\mu_h) \star \bm{U}(\{\underline{n}\},s,\mu_h,\mu_s) \notag\\
   &= \bm{\mathcal{H}}(\{\underline{n}\},s,\mu_h) + \int_{\mu_s}^{\mu_h}\!\frac{d\mu_1}{\mu_1}\,
    \bm{\mathcal{H}}(\{\underline{n}\},s,\mu_h) \star \bm{\Gamma}^H(\{\underline{n}\},s,\mu_1) \\[-1mm]
   &\quad + \int_{\mu_s}^{\mu_h}\!\frac{d\mu_1}{\mu_1} \int_{\mu_s}^{\mu_1}\!\frac{d\mu_2}{\mu_2}\,
    \bm{\mathcal{H}}(\{\underline{n}\},s,\mu_h) \star \bm{\Gamma}^H(\{\underline{n}\},s,\mu_1) 
    \star \bm{\Gamma}^H(\{\underline{n}\},s,\mu_2) + \dots \,, \notag
\end{align}
where the anomalous-dimension matrices on the right-hand side are ordered in the direction of decreasing scale values, i.e.\ $\mu_1>\mu_2$ in the last line. The successive applications of $\bm{\Gamma}^H$ lead to color structures with increasing complexity. We use this solution to evolve the hard functions from a hard matching scale $\mu_h\sim Q$ to a low scale $\mu_s\sim Q_0$, at which they are combined with the initial conditions
\begin{equation}\label{eq:LOlowenergyME}
   \bm{\mathcal{W}}_m(\{\underline{n}\},Q_0,\xi_1,\xi_2,\mu_s) 
   = f_1(\xi_1,\mu_s)\,f_2(\xi_2,\mu_s)\,\bm{1} + \mathcal{O}(\alpha_s)
\end{equation}
for the low-energy matrix elements. Here $f_i(\xi_i,\mu_s)$ denote the parton distribution functions. 

Following~\cite{Becher:2021zkk,Becher:2023mtx}, we are interested in the resummation of large double-logarithmic corrections to the cross section (the SLLs) at leading order in RG-improved perturbation theory. For this purpose it is sufficient to use the lowest-order expressions for the hard functions at the scale $\mu_h$ and the low-energy matrix elements at the scale $\mu_s$, and to solve the path-ordered exponential using a consistent leading-order approximation for the anomalous dimension, i.e.\ two-loop order for all logarithmically-enhanced terms, and one-loop order for the remaining terms. In general, one can split the anomalous dimensions into soft and collinear parts, $\bm{\Gamma}^S$ and $\bm{\Gamma}^C$, where only the soft part will be relevant in the following.\footnote{The explicit form of the collinear operator $\bm{\Gamma}^C$ has been worked out in~\cite{Becher:2023mtx}.} 
Soft emissions leave the values of the parton momentum fractions unchanged, and hence the Mellin convolutions are trivial for the terms in $\bm{\Gamma}^S$. They will thus be omitted from now on. It is convenient to split up the soft anomalous dimension in three terms~\cite{Becher:2021zkk}, 
\begin{equation}\label{eq:anomalous_dimension_soft_part}
   \bm{\Gamma}^S 
   = \gamma_{\rm cusp}(\alpha_s) \left( \bm{\Gamma}^c \ln\frac{\mu^2}{\mu_h^2} + \bm{V}^G \right)
    + \frac{\alpha_s}{4\pi}\,\overline{\bm{\Gamma}} + \mathcal{O}(\alpha_s^2) \,,
\end{equation}
where $\gamma_{\rm cusp}(\alpha_s)$ is the light-like cusp anomalous dimension~\cite{Korchemskaya:1992je}. The explicit expressions for the collinear-emission operator $\bm{\Gamma}^c$ and the Glauber operator $\bm{V}^G$ are~\cite{Becher:2021zkk}\footnote{Compared to this reference, we have removed a factor $\gamma_0=4$ from the definitions of the operators $\bm{\Gamma}^c$ and $\bm{V}^G$ and absorbed it into the cusp anomalous dimension.}
\begin{equation}\label{eq:VGGammac}
\begin{aligned} 
    \bm{\Gamma}^c 
    &= \sum_{i=1,2}\,\big[ C_i\,\bm{1} - \bm{T}_{i,L}\circ\bm{T}_{i,R}\,\delta(n_{k}-n_i) \big] \,, \\
    \bm{V}^G 
    &= - 2i\pi\,\big( \bm{T}_{1,L}\cdot\bm{T}_{2,L} - \bm{T}_{1,R}\cdot\bm{T}_{2,R} \big) \,.
\end{aligned}
\end{equation}
They account for (virtual or real) collinear emissions from one of the two initial-state partons, or Glauber-gluon exchanges between the two initial-state partons, respectively. The operator $\bm{V}^G$ is diagonal in multiplicity space, while $\bm{\Gamma}^c$ is an upper bi-diagonal matrix, and both only involve the color generators of the initial-state partons ($i=1,2$). We use the color-space formalism \cite{Catani:1996vz}, where $\bm{T}_i$ denotes a color generator acting on particle $i$, and $\bm{T}_{i}\cdot\bm{T}_{j}=\sum_a \bm{T}_{i}^a\,\bm{T}_{j}^a$. Moreover, $C_i\in\{C_F,C_A\}$ is the eigenvalue of the quadratic Casimir operator of $SU(N_c)$ in the fundamental or adjoint representation. The color matrices $\bm{T}_{i,L}$ ($\bm{T}_{i,R}$) act on the (conjugate) scattering amplitude and multiply the hard function from the left (right). The notation is different for the real-emission term in $\bm{\Gamma}^c$, which maps an amplitude with $m$ partons and associated color indices onto an amplitude with $(m+1)$ partons. The symbol $\circ$ indicates the extension of the color space, which now includes the emitted gluon with direction $n_k$. Explicitly, we have~\cite{Becher:2021zkk}
\begin{equation}
   \bm{\mathcal{H}}_m\,\bm{T}_{i,L}\circ\bm{T}_{j,R} 
   = \bm{T}_i^{a_k}\,\bm{\mathcal{H}}_m\,\bm{T}_j^{b_k} \,,
\end{equation}
where $a_k$ and $b_k$ are the color indices of the emitted gluon in the amplitude and the conjugate amplitude, respectively. In contrast to the virtual case, these color indices cannot immediately be contracted,  because later emissions can attach to the new gluon.

The emission operator $\overline{\bm{\Gamma}}$ in~\eqref{eq:anomalous_dimension_soft_part} accounts for (virtual or real) gluon emissions away from the directions of the jets. Its explicit expression reads~\cite{Becher:2021zkk,Becher:2023mtx}
\begin{equation}
   \overline{\bm{\Gamma}} 
   = 2 \sum_{(ij)} \left( \bm{T}_{i,L}\cdot\bm{T}_{j,L} + \bm{T}_{i,R}\cdot\bm{T}_{j,R} \right) 
    \int\frac{d\Omega(n_k)}{4\pi}\,\overline{W}_{ij}^k - 4 \sum_{(ij)} \bm{T}_{i,L}\circ\bm{T}_{j,R}\,
    \overline{W}_{ij}^{k}\,\Theta_{\rm hard}(n_{k}) \,,
\end{equation}
where the sum over $(ij)$ includes all unordered pairs of parton indices with $i\ne j$. This operator contains color generators for all partons in the process. The subtracted soft dipole (with collinear singularities removed) is defined as
\begin{equation}\label{eq:subtractedDipole}
   \overline{W}_{ij}^k 
   \equiv W_{ij}^k - \frac{1}{n_i\cdot n_k}\,\delta(n_i- n_k) - \frac{1}{n_j\cdot n_k}\,\delta(n_j- n_k) \,; \qquad
   W_{ij}^k = \frac{n_i\cdot n_j}{n_i\cdot n_k\,n_j\cdot n_k} \,.
\end{equation}
It is understood that the angular $\delta$-distributions act only on the test function, not on the coefficients multiplying them. The hard gluons in the real-emission term are restricted to lie inside the jet region, as indicated by the constraint $\Theta_{\rm hard}(n_{k})$, while the virtual corrections are unrestricted. 

The three operators in~\eqref{eq:anomalous_dimension_soft_part} satisfy the identities~\cite{Becher:2021zkk}
\begin{equation}\label{eq:prop_GammaC_GammaBar_VG}
    [\spac\bm{\Gamma}^c,\overline{\bm{\Gamma}}\spac] = 0 \,, \qquad
    \big\langle \bm{\mathcal{H}}\,\bm{\Gamma}^c\otimes\bm{1} \big\rangle = 0 \,, \qquad
    \big\langle \bm{\mathcal{H}}\,\bm{V}^G\otimes\bm{1} \big\rangle = 0 \,,
\end{equation}
where $\bm{\mathcal{H}}$ can be an arbitrary hard function. It follows that, if at least one particle in the initial state is color neutral and hence the Glauber operator $\bm{V}^G$ vanishes, the cross section does not receive double-logarithmic contributions. The SLLs in QCD cross sections are thus obtained from terms involving a single emission operator $\overline{\bm{\Gamma}}$, two Glauber phases (to obtain a real cross section), and an arbitrary number of insertions of the collinear-emission operator $\bm{\Gamma}^c$, leading to color traces of the form~\cite{Becher:2021zkk}
\begin{equation}\label{eq:colortraces}
   C_{rn} = \big\langle \bm{\mathcal{H}}_{2\to M} \left( \bm{\Gamma}^c \right)^r
    \bm{V}^G \left( \bm{\Gamma}^c \right)^{n-r} \bm{V}^G\,\overline{\bm{\Gamma}}\otimes\bm{1} \big\rangle \,,
\end{equation}
where $0\le r\le n$. The cyclicity of the color trace implies that the virtual- and real-emission contributions in the operator $\overline{\bm{\Gamma}}\otimes\bm{1}$ add up to produce the angular integral
\begin{equation}
   \int\frac{d\Omega(n_k)}{4\pi}\,\overline{W}_{ij}^k \left[ 1 - \Theta_{\rm hard}(n_{k}) \right] 
   = \int\frac{d\Omega(n_k)}{4\pi}\,W_{ij}^k\,\Theta_{\rm veto}(n_{k}) \,,
\end{equation}
corresponding to an emission into the veto region. At least one such soft emission is required to obtain a sensitivity to the veto scale $Q_0$, without which there would be no large logarithms in the cross section. Note that on the right-hand side of this relation the subtraction terms in~\eqref{eq:subtractedDipole} can be omitted, because the direction $n_k$ of the emitted gluon is not collinear with $n_1$ or $n_2$.

The color traces $C_{rn}$ are associated with terms contributing at $\mathcal{O}(\alpha_s^{n+3} L^{2n+3})$ in perturbation theory. Performing the relevant scale integrals in \eqref{eq:Uexp} using a fixed coupling $\alpha_s(\bar\mu)$ one obtains the result \eqref{eq:SLLgeneric}. The color traces associated with higher-order terms in the Glauber series are generalizations of the traces shown in~\eqref{eq:colortraces}, with additional insertions of Glauber operators $\bm{V}^G$ intertwined with insertions of powers of the collinear-emission operator $\bm{\Gamma}^c$, see relation~(3.1) in \cite{Boer:2023jsy}.

\section{Resummation in RG-improved perturbation theory}
\label{sec:Uoperator}

In previous work~\cite{Becher:2021zkk,Becher:2023mtx}, the evolution operator \eqref{eq:path_ordered_exponential} was expanded in a power series, see~\eqref{eq:Uexp}, and it was shown that at a given order in perturbation theory the terms with the largest number of double-logarithmic contributions (from insertions of $\bm{\Gamma}^c$) are associated with the color traces $C_{rn}$ in~\eqref{eq:colortraces}. For fixed values of $r$ and $n$, the relevant scale integrals $I_{rn}$ were evaluated, and the series of SLLs was then expressed as a double sum $\sum_{n=0}^\infty\sum_{r=0}^n I_{rn}\spac C_{rn}$. Using a fixed coupling $\alpha_s(\bar\mu)$ in the evaluation of the integrals $I_{rn}$, the double sum was expressed in closed form in terms of Kamp\'e de F\'eriet functions. However, the all-order resummation with a running coupling has not yet been achieved. 

In the present work, we develop an alternative resummation approach, in which all double-logarithmic corrections are exponentiated from the beginning. Starting from the series expansion~\eqref{eq:Uexp}, and noting that with the initial conditions~\eqref{eq:LOlowenergyME} for the low-energy matrix elements a non-zero color trace can only be obtained if the two right-most insertions of $\bm{\Gamma}^H$ are proportional to $\bm{V}^G\,\overline{\bm{\Gamma}}$, one can show that the infinite series of the SLLs is generated by the evolution operator
\begin{equation}\label{eq:Usll}
\begin{aligned}
   \bm{U}_{\rm SLL}(\{\underline{n}\},\mu_h,\mu_s) 
   &= \int_{\mu_s}^{\mu_h}\!\frac{d\mu_1}{\mu_1} \int_{\mu_s}^{\mu_1}\!\frac{d\mu_2}{\mu_2}
    \int_{\mu_s}^{\mu_2}\!\frac{d\mu_3}{\mu_3} \\
   &\quad\times \bm{U}_c(\mu_h,\mu_1)\,\gamma_{\rm cusp}\big(\alpha_s(\mu_1)\big)\,\bm{V}^G\,
    \bm{U}_c(\mu_1,\mu_2)\,\gamma_{\rm cusp}\big(\alpha_s(\mu_2)\big)\,\bm{V}^G\,
    \frac{\alpha_s(\mu_3)}{4\pi}\,\overline{\bm{\Gamma}} \,,
\end{aligned}
\end{equation}
where we have defined the generalized Sudakov operator
\begin{equation}\label{eq:Ucmuimuj}
   \bm{U}_c(\mu_i,\mu_j) 
   = \exp\bigg[ \bm{\Gamma}^c\!\int_{\mu_j}^{\mu_i}\!\frac{d\mu}{\mu}\,\gamma_{\rm cusp}\big(\alpha_s(\mu)\big) \ln\frac{\mu^2}{\mu_h^2} \bigg] \,.
\end{equation}
No path-ordering is required in this expression, since the matrix structure in the exponent is scale-independent. Expression~\eqref{eq:Usll} can be generalized straightforwardly to the case of additional Glauber-operator insertions, where for each factor of $\bm{V}^G$ one encounters a new scale integral. We define 
\begin{equation}\label{eq:USLLndef}
   \bm{U}_{\rm SLL}^{(l)}(\{\underline{n}\},\mu_h,\mu_s) 
   = \int_{\mu_s}^{\mu_h}\!\frac{d\mu_1}{\mu_1}\,\ldots\int_{\mu_s}^{\mu_l}\!\frac{d\mu_{l+1}}{\mu_{l+1}}
    \left[ \prod_{i=1}^l\,\bm{U}_c(\mu_{i-1},\mu_i)\,\gamma_{\rm cusp}\big(\alpha_s(\mu_i)\big)\,\bm{V}^G \right] 
    \frac{\alpha_s(\mu_{l+1})}{4\pi}\,\overline{\bm{\Gamma}} \,,
\end{equation}
where $\mu_0\equiv\mu_h$, and the terms in the product are ordered from left to right according to increasing values of $i$. Note that $\overline{\bm{\Gamma}}$ carries an implicit dependence on the directions $\{\underline{n}\}$ of the final-state particles. With this definition, the Glauber series can be written as
\begin{equation}\label{eq:Glauber_series}
   \bm{U}_{\rm SLL+G}(\{\underline{n}\},\mu_h,\mu_s) 
   = \sum_{l=1}^\infty\,\bm{U}_{\rm SLL}^{(l)}(\{\underline{n}\},\mu_h,\mu_s) \,.
\end{equation}

We note that the structure in~\eqref{eq:USLLndef} resembles the partially factorized operator series studied in~\cite{Forshaw:2019ver,Forshaw:2021fxs} as well as earlier papers on SLLs~\cite{Forshaw:2006fk,Forshaw:2008cq,Keates:2009dn}.
The novelty of the approach developed here and in~\cite{Becher:2023mtx,Boer:2023jsy,Boer:2023ljq} is that it allows for analytical and numerical resummations of the Glauber series for gap-between-jet observables to all orders in perturbation theory, a task previously thought to be impossible.

Since QCD Born-level hard functions are real, the sum over $l$ can be restricted to even numbers $l=2\ell$ in this case, where $\ell$ denotes the number of Glauber-operator pairs. Relation~\eqref{eq:USLLndef} then provides an alternative way in which to reproduce the results obtained in~\cite{Boer:2023jsy,Boer:2023ljq}. However, it has been pointed out in~\cite{Forshaw:2012bi} that hard functions for scattering processes involving the massive electroweak gauge bosons can be complex-valued even at tree level. In general, if the Born-level scattering amplitude is of the form $|\mathcal{M}_{2\to M}\rangle=|\mathcal{M}_{2\to M}^{(r)}\rangle+i\spac|\mathcal{M}_{2\to M}^{(i)}\rangle$, where the two terms have different color structures, then the hard function has the form (apart from phase-space integrations)
\begin{equation}
\begin{aligned}
   \bm{\mathcal{H}}_{2\to M}
   &\sim |\mathcal{M}_{2\to M}^{(r)}\rangle \langle\mathcal{M}_{2\to M}^{(r)}| 
    + |\mathcal{M}_{2\to M}^{(i)}\rangle \langle\mathcal{M}_{2\to M}^{(i)}| \\
   &\quad + i \left( |\mathcal{M}_{2\to M}^{(i)}\rangle \langle\mathcal{M}_{2\to M}^{(r)}| 
    - |\mathcal{M}_{2\to M}^{(r)}\rangle \langle\mathcal{M}_{2\to M}^{(i)}| \right) ,
\end{aligned}
\end{equation}
and contains a non-vanishing imaginary part. In this case, SLLs can be generated already with a single insertion of the Glauber operator $\bm{V}^G$ and, more generally, both even and odd values of $l$ in~\eqref{eq:Glauber_series} contribute to the Glauber series. In leading-logarithmic approximation, the contribution of the Glauber series to the cross section in~\eqref{eq:factorization_formula} is then obtained in the form
\begin{equation}
   \sigma_{2\to M}^{\rm SLL+G}(Q_0) = \sum_\text{partonic channels} \int\!d\xi_1 \int\!d\xi_2\,f_1(\xi_1,\mu_s)\,f_2(\xi_2,\mu_s) \, \hat\sigma_{2\to M}^{\rm SLL+G}(\xi_1,\xi_2,Q_0) \,,
\end{equation}
with partonic $2\to M$ scattering cross section
\begin{equation}
    \hat\sigma_{2\to M}^{\rm SLL+G}(\xi_1,\xi_2,Q_0) = \big\langle\bm{\mathcal{H}}_{2\to M}(\{\underline{n}\},s,\xi_1,\xi_2,\mu_h)\,
    \bm{U}_{\rm SLL+G}(\{\underline{n}\},\mu_h,\mu_s)\otimes\bm{1} \big\rangle \,,
\end{equation}
where $\mu_h\sim Q$ and $\mu_s\sim Q_0$.

{\em A priori}, the iterated insertions of the color operators $\bm{\Gamma}^c$ and $\bm{V}^G$ in the evolution operator generate increasingly more complicated color structures, because each insertion of the real-emission term in $\bm{\Gamma}^c$ in~\eqref{eq:VGGammac} adds a collinear gluon with its own color space. However, it has been shown in~\cite{Becher:2021zkk,Becher:2023mtx} that the infinite series of SLLs can be reduced to a {\em finite\/} set of basis operators $\{\bm{X}_i\}$ in color space. Remarkably, this statement remains true if higher-order terms in the Glauber series are included. The basis contains 5 operators for partonic scattering processes featuring quarks and/or anti-quarks in the initial state~\cite{Boer:2023jsy}. For gluon- and quark-gluon-initiated scattering processes, the basis contains 20 and 14 operators, respectively~\cite{Boer:2023ljq}. 

In this section, we discuss the simplest case of quark-initiated scattering in detail, for which the initial-state partons in the hard scattering are of the form $q_1 q_2$, $\bar q_1\bar q_2$, or $q_1\bar q_2$, where the flavors can be different. The cases of quark-gluon or gluon-gluon scattering have been considered in~\cite{Boer:2023ljq} and are discussed in Appendix~\ref{app:gluons}. For quark-initiated scattering, one finds that under the color trace the product $\bm{U}_{\rm SLL+G}(\{\underline{n}\},\mu_h,\mu_s)\otimes\bm{1}$ can be decomposed into the five color structures~\cite{Becher:2021zkk,Boer:2023jsy}
\begin{equation}\label{eq:Xbasis}
\begin{aligned}
   \bm{X}_1 &= \sum_{j>2} J_j\,if^{abc}\,\bm{T}_1^a\spac\bm{T}_2^b\spac\bm{T}_j^c \,, 
    &\bm{X}_4 &= \frac{1}{N_c}\,J_{12}\,\bm{T}_1\cdot\bm{T}_2 \,, \\
   \bm{X}_2 &= \sum_{j>2} J_j\,(\sigma_1-\sigma_2)\,d^{abc}\,\bm{T}_1^a\spac\bm{T}_2^b\spac\bm{T}_j^c \,, \qquad
    &\bm{X}_5 &= J_{12}\,\bm{1} \,, \\
   \bm{X}_3 &= \frac{1}{N_c} \sum_{j>2} J_j \left( \bm{T}_1 - \bm{T}_2\right)\cdot \bm{T}_j \,, &&
\end{aligned}
\end{equation}
times coefficients accounting for the dependence on the matching scales $\mu_h$ and $\mu_s$. Here $\sigma_i=1$ for an initial-state anti-quark and $\sigma_i=-1$ for an initial-state quark. The sum over $j$ in the first three structures extends over the final-state partons of the Born process. We have defined the angular integrals
\begin{equation}\label{eq:Jints}
\begin{aligned}
   J_j &= \int\frac{d\Omega(n_k)}{4\pi} \left( W_{1j}^k - W_{2j}^k \right) \Theta_{\rm veto}(n_k) \,, \\
   J_{12} &= \int\frac{d\Omega(n_k)}{4\pi}\,W_{12}^k\,\Theta_{\rm veto}(n_k) \,,
\end{aligned}
\end{equation}
which contain the information about the directions of the hard partons in the process. These integrals describe the emission of a soft gluon (emitted from particles~1, 2, or $j$) into the region outside the jets, in which a veto on energetic radiation is imposed. Note that compared with the conventions used in~\cite{Becher:2021zkk,Boer:2023jsy}, we have included a factor $1/N_c$ in the definitions of $\bm{X}_3$ and $\bm{X}_4$. This ensures that the contributions of all structures to a given partonic scattering process scale at most as $\mathcal{O}(1)$ in the large-$N_c$ limit, see Table~\ref{tab:ME_qq}.

\begin{table}
\centering
\begin{tabular}{|l||c|c|c|c|c|c|}
\hline
 & $q\bar q\to 0$ & $q\bar q\to g$ & $qq' \to qq'$ & $q\bar q \to q'\bar q'$ & $q\bar q'\to q \bar q'$
 & $q\bar q \to gg$ \\ \hline\hline
$\bm{X}_1$ & -- & $0$ & $0$ & $0$ & $0$ & $0$ \\[2mm]
$\bm{X}_2$ & -- & $0$ & $0$ & $\mathcal{O}(1)$ & $\mathcal{O}(1)$ & $0$ \\[2mm]
$\bm{X}_3$ & -- & $0$ & $\mathcal{O}(1)$ & $\mathcal{O}(1)$ & $\mathcal{O}(1/N_c^2)$ & $\mathcal{O}(1)$ \\[2mm]
$\bm{X}_4$ & $\mathcal{O}(1)$ & $\mathcal{O}(1/N_c^2)$ & $\mathcal{O}(1/N_c^2)$ & $\mathcal{O}(1/N_c^2)$
 & $\mathcal{O}(1)$ & $\mathcal{O}(1/N_c^4)$ \\[2mm]
$\bm{X}_5$ & $\mathcal{O}(1)$ & $\mathcal{O}(1)$ & $\mathcal{O}(1)$ & $\mathcal{O}(1)$ & $\mathcal{O}(1)$
 & $\mathcal{O}(1)$ \\ \hline
\end{tabular}
\caption{Large-$N_c$ counting of the matrix elements $\langle \bm{\mathcal{H}}_{2\to M}\,\bm{X}_i\rangle$ for quark-initiated partonic scattering processes, where quarks $q$ and $q'$ have different flavors.}
\label{tab:ME_qq}
\end{table}

We now define matrix representations of the collinear-emission operator $\bm{\Gamma}^c$ and the Glauber operator $\bm{V}^G$ on the space of basis structures, such that\footnote{Compared to~\cite{Boer:2023ljq} we define these matrices with transposed indices, because this preserves the order of matrix products.}
\begin{equation}\label{eq:matrix_reps}
\begin{aligned}
   \big\langle \bm{\mathcal{H}}\,\bm{\Gamma}^c\bm{X}_i \big\rangle
   &= \sum_{\tilde{\imath}} \big\langle \bm{\mathcal{H}}\,\bm{X}_{\tilde{\imath}} \big\rangle\,N_c 
    \left( \GGammac \right)_{\tilde{\imath} i} \spac , \\
   \big\langle \bm{\mathcal{H}}\,\bm{V}^G\bm{X}_i \big\rangle
   &= \sum_{\tilde{\imath}} \big\langle \bm{\mathcal{H}}\,\bm{X}_{\tilde{\imath}} \big\rangle\spac\,
    i\pi\spac N_c \left( \mathbbm{V}^G \right)_{\tilde{\imath} i} \spac ,
\end{aligned}
\end{equation}
where $\bm{\mathcal{\bm{\mathcal{H}}}}$ can be a generic hard function. Next, we define the row vectors $\bm{X}^T\hspace{-0.2em}=(\bm{X}_1,\dots,\bm{X}_5)$ and $\varsigma^T=(1,0,0,0,0)$. Under the color trace, the operator $\bm{V}^G\,\overline{\bm{\Gamma}}\otimes\bm{1}$ maps onto~\cite{Becher:2023mtx}
\begin{equation}\label{eq:firstGlauber}
   \bm{V}^G\,\overline{\bm{\Gamma}}\otimes\bm{1} \to 16i\pi\spac\bm{X}_1
   = 16i\pi\spac\bm{X}^T\spac\varsigma \,.
\end{equation}
The absence of $N_c^2$ on the right-hand side of this relation is responsible for the suppression of the SLLs, and in fact of all terms in the Glauber series, by a factor $1/N_c^2$ in the large-$N_c$ limit. It now follows that
\begin{equation}\label{eq:sigmaSLLG}
    \hat\sigma_{2\to M}^{\rm SLL+G}(Q_0) = \sum_{l=1}^\infty \big\langle\bm{\mathcal{H}}_{2\to M}(\mu_h)\,\bm{X}^T \big\rangle\,
    \mathbbm{U}_{\rm SLL}^{(l)}(\mu_h,\mu_s)\,\varsigma \,,
\end{equation}
where
\begin{equation}\label{eq:Usll_matrix_rep}
\begin{aligned}   
   \mathbbm{U}_{\rm SLL}^{(l)}(\mu_h,\mu_s) 
   &= 16 \left( i\pi \right)^l N_c^{l-1} \int_{\mu_s}^{\mu_h}\!\frac{d\mu_1}{\mu_1}\,\ldots
    \int_{\mu_s}^{\mu_l}\!\frac{d\mu_{l+1}}{\mu_{l+1}}\,\mathbbm{U}_c(\mu_h,\mu_1) \\
    &\quad\times \left[ \prod_{i=1}^{l-1}\,\gamma_{\rm cusp}\big(\alpha_s(\mu_i)\big)\,\mathbbm{V}^G\,
     \mathbbm{U}_c(\mu_i,\mu_{i+1}) \right] \gamma_{\rm cusp}\big(\alpha_s(\mu_l)\big)\, 
     \frac{\alpha_s(\mu_{l+1})}{4\pi} \,,
\end{aligned}
\end{equation}
and
\begin{equation}\label{eq:Ucexp}
   \mathbbm{U}_c(\mu_i,\mu_j) 
   = \exp\bigg[ N_c\,\GGammac\!\int_{\mu_j}^{\mu_i}\!\frac{d\mu}{\mu}\,\gamma_{\rm cusp}\big(\alpha_s(\mu)\big)
    \ln\frac{\mu^2}{\mu_h^2} \bigg] \,.
\end{equation}
The explicit forms of the matrices $\GGammac$ and $\mathbbm{V}^G$ for quark-initiated scattering read~\cite{Boer:2023jsy} 
\begin{equation}\label{eq:GGammacVG}
   \GGammac
   = \begin{pmatrix}
    1~~ & 0 & 0 & 0~~ & 0 \\
    0~~ & 1 & 0 & 0~~ & 0 \\
    0~~ & 0 & \frac12 & 0~~ & 0 \\
    0~~ & 0 & -1 & 1~~ & 0 \\
    0~~ & 0 & -\frac{C_F}{N_c} & 0~~ & 0
   \end{pmatrix} , \qquad
   \mathbbm{V}^G 
   = \begin{pmatrix}
    0 & -2\spac\delta_{q\bar q}\,\frac{N_c^2-4}{N_c^2} & ~\frac{4}{N_c^2} & ~~0 & ~~0 \\
    - \frac12 & 0 & ~0 & ~~0 & ~~0 \\
    1 & 0 & ~0 & ~~0 & ~~0 \\
    0 & 0 & ~0 & ~~0 & ~~0 \\
    0 & 0 & ~0 & ~~0 & ~~0
    \end{pmatrix} ,
\end{equation}
where $\delta_{q\bar q}\equiv\frac14\spac(\sigma_1-\sigma_2)^2$ equals~1 for the $q\bar q'$ initial states, and~0 for $q q'$ or $\bar q\bar q'$ initial states. The matrix exponential~\eqref{eq:Ucexp} now takes the form 
\begin{equation}\label{eq:Ucdiag}
   \mathbbm{U}_c(\mu_i,\mu_j) 
   = \begin{pmatrix}
    U_c(1; \mu_i,\mu_j) & 0 & 0 & 0 & ~0 \\
    0 & U_c(1; \mu_i,\mu_j) & 0 & 0 & ~0 \\
    0 & 0 & U_c(\textstyle{\frac12}; \mu_i,\mu_j) & 0 & ~0 \\
    0 & 0 & 2 \left[ U_c(\textstyle{\frac12}; \mu_i,\mu_j) - U_c(1; \mu_i,\mu_j) \right]
     & ~U_c(1; \mu_i,\mu_j)~ & ~0 \\
    0 & 0 & \frac{2\spac C_F}{N_c} \left[ 1 - U_c(\textstyle{\frac12}; \mu_i,\mu_j) \right] & 0 & ~1
    \end{pmatrix} ,
\end{equation}
with
\begin{equation} \label{eq:scalar_evolution_function}
   U_c(v; \mu_i,\mu_j) 
   = \exp\bigg[ v\spac N_c 
    \int_{\mu_j}^{\mu_i}\!\frac{d\mu}{\mu}\,\gamma_{\rm cusp}\big(\alpha_s(\mu)\big)
    \ln\frac{\mu^2}{\mu_h^2} \bigg] \,.
\end{equation}
For $v\ge 0$, this function satisfies $0<U_c(v; \mu_i,\mu_j)\le 1$, where the value~1 is obtained only for $\mu_i=\mu_j$ or $v=0$. In general, $v$ is equal to one of the eigenvalues of $\GGammac$.

It is instructive to explore the matrix structure of the result \eqref{eq:Usll_matrix_rep} in more detail, using that
\begin{equation}\label{eq:VGUcmat}
   \mathbbm{V}^G\,\mathbbm{U}_c(\mu_i,\mu_j)
   = \begin{pmatrix}
    0 & - 2\spac\delta_{q\bar q}\,\frac{N_c^2-4}{N_c^2}\,U_c(1; \mu_i,\mu_j)
     & ~\frac{4}{N_c^2}\,U_c(\textstyle{\frac12}; \mu_i,\mu_j)~ & 0~ & ~~0 \\
    - \frac12\,U_c(1; \mu_i,\mu_j) & 0 & 0 & 0~ & ~~0 \\
    U_c(1; \mu_i,\mu_j) & 0 & 0 & 0~ & ~~0 \\
    0 & 0 & 0 & 0~ & ~~0 \\
    0 & 0 & 0 & 0~ & ~~0
    \end{pmatrix} . 
\end{equation}
The eigenvalue~0 does not appear in this expression and, therefore, can only contribute through the leftmost factor $\mathbbm{U}_c(\mu_h,\mu_1)$ in~\eqref{eq:Usll_matrix_rep}. The multiplication with the vector $\varsigma$ in~\eqref{eq:sigmaSLLG} projects out the first column of the product of $(l-1)$ such matrices. It follows that for odd values of $l$, only the first component of the resulting vector is non-zero, while for even values of $l$, the first component vanishes but the remaining four components are non-zero. Using the fact that the vector $\varsigma$ is an eigenvector of $\GGammac$ with eigenvalue~1, which is true irrespective of the nature of the initial-state partons~\cite{Becher:2021zkk}, the rightmost factor $\mathbbm{U}_c(\mu_{l-1},\mu_l)$ always generates $U_c(1;\mu_{l-1},\mu_l)$. Note also that the (1,\spac 3) entry of the product $\mathbbm{V}^G\,\mathbbm{U}_c(\mu_i,\mu_j)$, which contains the only contribution corresponding to the eigenvalue~$\frac12$, vanishes in the large-$N_c$ limit. This fact greatly simplifies the treatment of higher-order terms in the Glauber series in the large-$N_c$ approximation~\cite{Boer:2024xzy}.

In the above result, all double-logarithmic effects are resummed into the generalized Sudakov factors $U_c(v; \mu_i,\mu_j)$ with eigenvalues $v\in\{\frac12,1\}$. It is straightforward to evaluate these factors at leading order in RG-improved perturbation theory. To this end, we change variables from $\mu$ to the running coupling $\alpha_s(\mu)$ via $d\alpha_s(\mu)/d\ln\mu=\beta\big(\alpha_s(\mu)\big)$ and use the perturbative expansions of the QCD $\beta$-function and the cusp anomalous dimension,
\begin{equation}\label{eq:pertexp}
   \beta(\alpha_s) 
    = - 2\alpha_s \sum_{n=0}^\infty\,\beta_n \left( \frac{\alpha_s}{4\pi} \right)^{n+1} \,, \qquad   
   \gamma_{\rm cusp}(\alpha_s) 
    = \sum_{n=0}^\infty\,\gamma_n \left( \frac{\alpha_s}{4\pi} \right)^{n+1} ,
\end{equation}
with expansion coefficients given in Appendix~\ref{app:Sudakov}. We find
\begin{equation}\label{eq:UcSuda}
\begin{aligned}
   U_c(v; \mu_i,\mu_j)
   &= \exp\Bigg\{ \frac{\gamma_0\spac v N_c}{2\beta_0^2}\,\Bigg[ 
    \frac{4\pi}{\alpha_s(\mu_h)} \left( \frac{1}{x_i} - \frac{1}{x_j} - \ln\frac{x_j}{x_i} \right) \\
   &\hspace{1.5cm}
    + \left( \frac{\gamma_1}{\gamma_0} - \frac{\beta_1}{\beta_0} \right) 
    \left( x_i - x_j + \ln\frac{x_j}{x_i} \right)
    + \frac{\beta_1}{2\beta_0} \left( \ln^2 x_j - \ln^2 x_i \right) \Bigg] \Bigg\} \,, 
\end{aligned}
\end{equation}
where $x_i\equiv\alpha_s(\mu_i)/\alpha_s(\mu_h)$. In the exponent, it is important to keep the two-loop approximations for the cusp anomalous dimension and the $\beta$-function. For the special case $\mu_i=\mu_h$ and $v=1$, this result reduces to the well-known expression for the Sudakov exponent $S(\mu_h,\mu_j)$ encountered in applications of soft-collinear effective theory~\cite{Neubert:2004dd}. We note the useful identities
\begin{equation}\label{eq:simplify}
   U_c(v; \mu_i,\mu_j)\,U_c(v; \mu_j,\mu_k) = U_c(v; \mu_i,\mu_k) \,, \qquad
   U_c(0; \mu_i,\mu_j) = 1 \,,
\end{equation}
which can be used to simplify products of functions $U_c(v; \mu_i,\mu_j)$. In Appendix~\ref{app:Sudakov} we also show how the expression~\eqref{eq:UcSuda} must be generalized if the lower scale lies below the top-quark threshold ($\mu_j<\mu_t$). 

In~\eqref{eq:UcSuda} we have expressed the generalized Sudakov factor $U_c(v; \mu_i,\mu_j)$ as a function of $x_i$ and $x_j$. The remaining integrals in~\eqref{eq:Usll_matrix_rep} can also be recast as integrals over the variables $x_i$. Inverting the order of the integrals, we obtain
\begin{equation}\label{eq:masterints}
\begin{aligned}
   \mathbbm{U}_{\rm SLL}^{(l)}(\mu_h,\mu_s) 
   &= \left( i\pi \right)^l N_c^{l-1}\,\frac{2^{l+3}}{\beta_0^{l+1}} 
    \int_1^{x_s}\!\frac{dx_l}{x_l}\,\ln\frac{x_s}{x_l} \int_1^{x_l}\!\frac{dx_{l-1}}{x_{l-1}}\,\ldots
    \int_1^{x_2}\!\frac{dx_1}{x_1} \\
   &\quad\times \mathbbm{U}_c(\mu_h,\mu_1) \left[ \prod_{i=1}^{l-1}\,\mathbbm{V}^G\,\mathbbm{U}_c(\mu_i,\mu_{i+1}) \right] ,
\end{aligned}
\end{equation}
where we have already performed the integral over $x_{l+1}$ and used the one-loop approximation for the cusp anomalous dimension in the Glauber terms (with $\gamma_0=4$). This formula accomplishes the resummation of the infinite series of terms involving $l$ insertions of Glauber operators at leading order in RG-improved perturbation theory. Introducing the shorthand notation
\begin{equation}\label{eq:UcChains}
   U_c(v^{(1)},\dots,v^{(l)}; \mu_h,\mu_1,\dots,\mu_l)
   \equiv U_c(v^{(1)}; \mu_h,\mu_1)\,U_c(v^{(2)}; \mu_1,\mu_2) \dots U_c(v^{(l)}; \mu_{l-1},\mu_l) \,,
\end{equation}
we obtain for the first two terms
\begin{equation}\label{eq:USLL1}
   \mathbbm{U}_{\rm SLL}^{(1)}(\mu_h,\mu_s)\,\varsigma
   = \frac{16i\pi}{\beta_0^2} \int_1^{x_s}\!\frac{dx_1}{x_1}\,\ln\frac{x_s}{x_1}\,
    U_c(1; \mu_h,\mu_1)\,\varsigma \,,
\end{equation}
where the right-hand side is proportional to the vector $\varsigma$, i.e.\ only its first component is non-zero, and 
\begin{equation}\label{eq:USLL2}
\begin{aligned}
   \mathbbm{U}_{\rm SLL}^{(2)}(\mu_h,\mu_s)\,\varsigma 
   &= - \frac{32\pi^2}{\beta_0^3}\,N_c \int_1^{x_s}\!\frac{dx_2}{x_2}\,\ln\frac{x_s}{x_2} 
    \int_1^{x_2}\!\frac{dx_1}{x_1} \\
   &\quad\times 
    \begin{pmatrix}
     0 \\
     - \frac12\,U_c(1; \mu_h,\mu_2) \\
     U_c(\textstyle{\frac12},1; \mu_h,\mu_1,\mu_2) \\
     2 \left[ U_c(\textstyle{\frac12},1; \mu_h,\mu_1,\mu_2) - U_c(1; \mu_h,\mu_2) \right] \\
     \frac{2\spac C_F}{N_c} \left[ U_c(1; \mu_1,\mu_2) - U_c(\textstyle{\frac12},1; \mu_h,\mu_1,\mu_2) \right]
    \end{pmatrix} .
\end{aligned}
\end{equation}
Here, we have used the identities
\begin{equation}
   U_c(1,1; \mu_h,\mu_1,\mu_2)
    = U_c(1; \mu_h,\mu_2) \,, \qquad
   U_c(0,1; \mu_h,\mu_1,\mu_2)
    = U_c(1; \mu_1,\mu_2) \,, 
\end{equation}
which follow from~\eqref{eq:simplify}. As explained above, the last eigenvalue always equals~1 and a zero eigenvalue can only appear in the first entry. Pressing on to higher $l$ values, we find
\begin{equation}\label{eq:USLL3}
\begin{aligned}
   \mathbbm{U}_{\rm SLL}^{(3)}(\mu_h,\mu_s)\,\varsigma
   &= - \frac{64\spac i\pi^3}{\beta_0^4}\,N_c^2 \int_1^{x_s}\!\frac{dx_3}{x_3}\,\ln\frac{x_s}{x_3} 
    \int_1^{x_3}\!\frac{dx_2}{x_2} \int_1^{x_2}\!\frac{dx_1}{x_1} \\
    &\quad\times \left[ K_{12}\,U_c(1;\mu_h,\mu_3) + \textstyle{\frac{4}{N_c^2}}\,
     U_c(1,\textstyle{\frac12},1; \mu_h,\mu_1,\mu_2,\mu_3) \right] \varsigma ,
\end{aligned}
\end{equation}
and 
\begin{equation}\label{eq:USLL4}
\begin{aligned}
   \mathbbm{U}_{\rm SLL}^{(4)}(\mu_h,\mu_s)\,\varsigma
   &= \frac{128\spac\pi^4}{\beta_0^5}\,N_c^3
    \int_1^{x_s}\!\frac{dx_4}{x_4}\,\ln\frac{x_s}{x_4} \int_1^{x_4}\!\frac{dx_3}{x_3} 
    \int_1^{x_3}\!\frac{dx_2}{x_2} \int_1^{x_2}\!\frac{dx_1}{x_1} \\
   &\quad\times
    \begin{pmatrix}
     0 \\[3mm]
     - \frac12 \left[ K_{12}\,U_c(1;\mu_h,\mu_4) + \textstyle{\frac{4}{N_c^2}}\,
      U_c(1,\textstyle{\frac12},1; \mu_h,\mu_2,\mu_3,\mu_4) \right] \\[4mm]
     K_{12}\,U_c(\frac12,1;\mu_h,\mu_1,\mu_4) + \textstyle{\frac{4}{N_c^2}}\,
      U_c(\textstyle{\frac12},1,\textstyle{\frac12},1; \mu_h,\mu_1,\mu_2,\mu_3,\mu_4) \\[4mm]
     2 \Big[ K_{12}\,U_c(\frac12,1;\mu_h,\mu_1,\mu_4) + \textstyle{\frac{4}{N_c^2}}\,
      U_c(\textstyle{\frac12},1,\textstyle{\frac12},1; \mu_h,\mu_1,\mu_2,\mu_3,\mu_4) \\
     - K_{12}\,U_c(1;\mu_h,\mu_4) - \textstyle{\frac{4}{N_c^2}}\,
      U_c(1,\textstyle{\frac12},1; \mu_h,\mu_2,\mu_3,\mu_4) \Big]
      \\[4mm]
     \frac{2\spac C_F}{N_c} \Big[ K_{12}\,U_c(1;\mu_1,\mu_4) + \textstyle{\frac{4}{N_c^2}}\, U_c(1,\textstyle{\frac12},1; \mu_1,\mu_2,\mu_3,\mu_4) \\
     \qquad- K_{12}\,U_c(\frac12,1;\mu_h,\mu_1,\mu_4) - \textstyle{\frac{4}{N_c^2}}\,
      U_c(\textstyle{\frac12},1,\textstyle{\frac12},1; \mu_h,\mu_1,\mu_2,\mu_3,\mu_4) \Big]     
    \end{pmatrix} ,
\end{aligned}
\end{equation}
where
\begin{equation}
   K_{12} \equiv (\sigma_1-\sigma_2)^2\,\frac{N_c^2-4}{4\spac N_c^2}
   = \frac{N_c^2-4}{N_c^2}\,\delta_{q\bar q} \,.
\end{equation}
For up to four insertions ($l\le 4$), the integrals over the $x_i$ variables can be performed numerically without much effort. Some results are presented in Section~\ref{sec:numerics}.

\section{Fixed-coupling results and asymptotic behavior}
\label{sec:asymptotics}

An important open challenge is to determine the asymptotic behavior of the resummed terms in the Glauber series in the limit where $\alpha_s\spac L_s\sim 1$ and hence $\alpha_s\spac L_s^2\gg 1$. Here, the variable $L_s=\ln(\mu_h/\mu_s)$ depends on the ratio of the hard and soft matching scales and reduces to the variable $L$ used in the introduction for the default scale choices $\mu_h=Q$ and $\mu_s=Q_0$. Adapting the definitions given in the text following relation~\eqref{eq:SLLgeneric} to more general scale choices, we define the variables
\begin{equation}
   w = \frac{N_c\,\alpha_s(\bar\mu)}{\pi}\,L_s^2 \,, \qquad
   w_\pi = \frac{N_c\,\alpha_s(\bar\mu)}{\pi}\,\pi^2 \,.
\end{equation}
For the series of the SLLs, corresponding to $l=2$ in~\eqref{eq:USLLndef}, it was shown in~\cite{Becher:2021zkk,Becher:2023mtx} that in the asymptotic limit where $w\gg 1$
\begin{equation}
   \bm{U}_{\rm SLL}(\{\underline{n}\},\mu_h,\mu_s)
   \sim \pi^2\spac N_c \left( \frac{\alpha_s(\bar\mu)\spac L_s}{\pi} \right)^3 \frac{\ln w}{w} 
   = \frac{\alpha_s(\bar\mu)\spac L_s}{\pi\spac N_c}\,w_\pi \ln w \,,
\end{equation}
but so far no corresponding estimates for subleading logarithmic corrections have been obtained. Our new formula~\eqref{eq:Usll_matrix_rep} provides a convenient framework for performing studies of the asymptotic behavior for $\alpha_s\spac L_s^2\gg 1$, since it resums all double-logarithmic corrections. 

To study the double-logarithmic asymptotics, it is sufficient to evaluate the evolution operators with a fixed coupling $\alpha_s\equiv\alpha_s(\bar\mu)$, since the scale dependence of the running coupling is a single-logarithmic effect. Using the one-loop approximation for the cusp anomalous dimension, we find from~\eqref{eq:Usll_matrix_rep}
\begin{equation}\label{eq:USLL_fixed_coupling_def}
\begin{aligned}
   \mathbbm{U}_{\rm SLL}^{(l)}(\mu_h,\mu_s) 
   &= 4 \left( i\pi \right)^l N_c^{l-1} \left( \frac{\alpha_s}{\pi} \right)^{l+1} 
    \int_0^{L_s}\!dL_1\,\ldots\int_{L_{l-1}}^{L_s}\!dL_l \left( L_s-L_l \right) \\
   &\quad\times \mathbbm{U}_c(\mu_h,\mu_1) \left[ \prod_{i=1}^{l-1}\,\mathbbm{V}^G\,\mathbbm{U}_c(\mu_i,\mu_{i+1}) \right] ,
\end{aligned}
\end{equation}
with $L_i\equiv\ln(\mu_h/\mu_i)\ge 0$, $\mathbbm{U}_c(\mu_{i-1},\mu_i)$ as given in \eqref{eq:Ucdiag}, and
\begin{equation}\label{eq:Sudakov}
   U_c(v; \mu_i,\mu_j) 
   \equiv \exp\left[ v\,\frac{N_c\spac\alpha_s}{\pi} \left( L_i^2 - L_j^2 \right) \right] .
\end{equation}

As in the previous section, we need to evaluate multi-dimensional integrals over the concatenations $U_c(v^{(1)},\dots,v^{(l)}; \mu_h,\mu_1,\dots,\mu_l)$ defined in~\eqref{eq:UcChains}, with $v^{(l)}=1$ and $v^{(i\ne l)}\in\{1,\frac12,0\}$, where a zero eigenvalue is only allowed for $v^{(1)}$. Inverting the order of the integrations, we define
\begin{equation}\label{eq:Sigmadef}
\begin{aligned}
   & \left( \frac{\alpha_s}{\pi} \right)^{l+1} 
    \int_0^{L_s}\!dL_l \left( L_s-L_l \right) \ldots\int_0^{L_2}\!dL_1\, 
    U_c(v^{(1)},\dots,v^{(l)}; \mu_h,\mu_1,\dots,\mu_l) \\
   &\equiv \frac{1}{(l+1)!} \left( \frac{\alpha_s}{\pi}\,L_s \right)^{l+1} 
    \Sigma(v^{(1)},\dots,v^{(l)};w) \,,
\end{aligned}
\end{equation}
where $\Sigma(v^{(1)},\dots,v^{(l)}; 0)=1$. In analogy with relations~\eqref{eq:USLL1},~\eqref{eq:USLL2} and~\eqref{eq:USLL3},~\eqref{eq:USLL4}, we then obtain for $l\le 4$
\begin{align}
   \mathbbm{U}_{\rm SLL}^{(1)}(\mu_h,\mu_s)\,\varsigma
   &= 2i\pi \left( \frac{\alpha_s}{\pi}\,L_s \right)^2 \Sigma(1;w)\,\varsigma \,, \\
   \mathbbm{U}_{\rm SLL}^{(2)}(\mu_h,\mu_s)\,\varsigma 
   &= - \frac{2\pi^2}{3}\,N_c \left( \frac{\alpha_s}{\pi}\,L_s \right)^3
    \begin{pmatrix}
     0 \\
     - \frac12\,\Sigma(1,1;w) \\
     \Sigma(\textstyle{\frac12},1;w) \\
     2 \left[ \Sigma(\textstyle{\frac12},1;w) - \Sigma(1,1;w) \right] \\
     \frac{2\spac C_F}{N_c} \left[ \Sigma(0,1;w) - \Sigma(\textstyle{\frac12},1;w) \right] \\
    \end{pmatrix} ,
    \label{eq:USLL2_fixed} \\
    \mathbbm{U}_{\rm SLL}^{(3)}(\mu_h,\mu_s)\,\varsigma
    &= - \frac{i\pi^3}{6}\,N_c^2 \left( \frac{\alpha_s}{\pi}\,L_s \right)^4
     \left[ K_{12}\,\Sigma(1,1,1;w) + \textstyle{\frac{4}{N_c^2}}\,
     \Sigma(1,\textstyle{\frac12},1;w) \right] \varsigma ,
     \label{eq:USLL3_fixed} \\ 
    \mathbbm{U}_{\rm SLL}^{(4)}(\mu_h,\mu_s)\,\varsigma 
    &= \frac{\pi^4}{30}\,N_c^3 \left( \frac{\alpha_s}{\pi}\,L_s \right)^5
     \begin{pmatrix}
        0 \\[2mm]
        - \frac12 \left[ K_{12}\,\Sigma(1,1,1,1;w) + \textstyle{\frac{4}{N_c^2}}\,
         \Sigma(1,1,\frac12,1;w) \right] \\[3.5mm]
        K_{12}\,\Sigma(\frac12,1,1,1;w) + \textstyle{\frac{4}{N_c^2}}\,
         \Sigma(\frac12,1,\frac12,1;w) \\[3.5mm]
        2 \Big[ K_{12}\,\Sigma(\frac12,1,1,1;w) + \textstyle{\frac{4}{N_c^2}}\,
         \Sigma(\frac12,1,\frac12,1;w) \\
        \quad - K_{12}\,\Sigma(1,1,1,1;w) - \textstyle{\frac{4}{N_c^2}}\,
         \Sigma(1,1,\frac12,1;w) \Big]
         \\[3.5mm]
        \frac{2\spac C_F}{N_c} \Big[ K_{12}\,\Sigma(0,1,1,1;w) + \textstyle{\frac{4}{N_c^2}}\, \Sigma(0,1,\frac12,1;w) \\
        \qquad - K_{12}\,\Sigma(\frac12,1,1,1;w) - \textstyle{\frac{4}{N_c^2}}\, 
         \Sigma(\frac12,1,\frac12,1;w) \Big]       
    \end{pmatrix} .
    \label{eq:USLL4_fixed} 
\end{align}
In the following, we collect explicit expressions for the relevant functions as well as their asymptotic expansions for $w\gg 1$.

\subsubsection*{\boldmath Case $l=1$}

We find
\begin{equation}
   \Sigma(1;w) 
   = \frac{\sqrt{\pi\spac w}\,\text{erf}\spac\big(\sqrt{w}\big)+e^{-w}-1}{w}
   = \frac{\sqrt{\pi}}{\sqrt{w}} - \frac{1}{w} + \mathcal{O}(e^{-w}) \,,
\end{equation}
where the last formula shows the complete asymptotic expansion for $w\gg 1$ up to exponentially small terms.

\subsubsection*{\boldmath Case $l=2$}

The object $\Sigma(v,1;w)$ is a Kamp\'e de F\'eriet function, for which several useful expressions have been derived in~\cite{Becher:2023mtx}. For the relevant values of $v$, we obtain
\begin{equation} \label{eq:sigma_l2}
\begin{aligned}
   \Sigma(1,1;w)
   &= \frac{3}{w} - \frac{3\sqrt{\pi}}{2\spac w^{3/2}}\,\text{erf}\spac\big(\sqrt{w}\big) \,, \\
   \Sigma(\textstyle{\frac12},1;w) 
   &= \frac{3\sqrt2}{2\spac w} \bigg[ 2 \ln(1+\sqrt2) 
    + 4i\pi\,T\bigg(\!\sqrt{2w},\frac{i}{\sqrt2}\bigg) \bigg] \\
   &\quad - \frac{3\sqrt{2\pi}}{2\spac w^{3/2}} \bigg[ \text{erf}\spac\bigg(\!{\sqrt{\frac{w}{2}}}\bigg) - e^{-w}\,\text{erfi}\spac\bigg(\!\sqrt{\frac{w}{2}}\bigg) \bigg] , \\    
   \Sigma(0,1;w) 
   &= {}_2F_2(1,1; 2,\textstyle{\frac52}; -w) \,.
\end{aligned}
\end{equation}
where 
\begin{equation}
   \text{erf}(z) = \frac{2}{\sqrt\pi} \int_0^z\!dx\,e^{-x^2} \,, \qquad
   \text{erfi}(z) = \frac{2}{\sqrt\pi} \int_0^z\!dx\,e^{x^2}
\end{equation}
are error functions, ${}_2F_2({a_1,a_2};{b_1,b_2};z)$ is a generalized hypergeometric function, and
\begin{equation}
   T(x,a) = \frac{1}{2\pi} \int_0^a\!dt\,\frac{e^{-\frac12 x^2\spac(1+t^2)}}{1+t^2}
\end{equation}
denotes the Owen $T$-function. To derive the result for $\Sigma(\textstyle{\frac12},1;w)$ from the corresponding expression given in~\cite{Becher:2023mtx}, we have used the identity
\begin{equation}
   T\bigg(ix\spac a,\frac{i}{a}-\varepsilon\bigg) + \frac14
   = T\big(x,ia) + \frac{i}{4}\,\text{erf}\spac\bigg(\frac{x}{\sqrt2}\bigg)\,
    \text{erfi}\spac\bigg(\frac{x\spac a}{\sqrt2}\bigg) \,,
\end{equation}
which holds for $0<a<1$. The $\varepsilon\to 0^+$ regulator is needed to regularize the pole at $t=i$. To prove this relation, one shows that the derivatives with respect to $x$ are identical on both sides, and that the relation holds for $x=0$.

The asymptotic behavior of these Kamp\'e de F\'eriet functions for $w\gg 1$ is given by
\begin{equation}\label{eq:l2asy}
\begin{aligned}
   \Sigma(1,1;w)
   &= \frac{3}{w} - \frac{3\sqrt{\pi}}{2\spac w^{3/2}} + \mathcal{O}(e^{-w}) \,, \\
   \Sigma(\textstyle{\frac12},1;w) 
   &= \frac{3\sqrt{2}\spac\ln(1+\sqrt{2})}{w} - \frac{3\sqrt{\pi}}{\sqrt{2}\,w^{3/2}} 
    + \mathcal{O}(e^{-\frac{w}{2}}) \,, \\
   \Sigma(0,1;w) 
   &= \frac32\,\frac{\ln(4w) + \gamma_E - 2}{w} + \frac{3}{4\spac w^2} + \mathcal{O}(w^{-3}) \,,
\end{aligned}
\end{equation}
where in the last case there exist higher-order power-suppressed terms in addition to exponentially small contributions.

\subsubsection*{\boldmath Case $l=3$}

In~\eqref{eq:USLL3_fixed} we only need the two functions
\begin{equation}
\begin{aligned}
   \Sigma(1,1,1;w) 
   &= 3\,\frac{\sqrt{\pi\spac w}\,\text{erf}\spac\big(\sqrt{w}\big) + 2\spac e^{-w} - 2}{w^2} \,, \\
   \Sigma(1,\textstyle{\frac12},1;w)
   &= 12 \int_0^1\!dz\,(1-z)\spac z^2\,e^{-z^2 w}\,{}_2F_2(1,1; \textstyle{\frac32},2; z^2\spac\frac{w}{2}) \,, 
\end{aligned}
\end{equation}
whose asymptotic expansions read
\begin{equation}
\begin{aligned}
   \Sigma(1,1,1;w) 
   &= \frac{3\sqrt{\pi}}{w^{3/2}} - \frac{6}{w^2} + \mathcal{O}(e^{-w}) \,, \\
   \Sigma(1,\textstyle{\frac12},1;w)
   &= \frac{6\sqrt{\pi}\,\ln 2}{w^{3/2}} - \frac{3\pi}{w^2} + \mathcal{O}(e^{-\frac{w}{2}}) \,.
\end{aligned}
\end{equation}
To obtain the second result, we have replaced the upper integration limit by infinity, using that the hypergeometric function behaves like ${}_2F_2(1,1;\textstyle{\frac32},2;y)\sim y^{-3/2}\,e^y$ for $y\to\infty$.

\subsubsection*{\boldmath Case $l=4$}

Relation~\eqref{eq:USLL4_fixed} involves six functions, four of which can be expressed in closed form. They are 
\begin{equation}
\begin{aligned}
   \Sigma(1,1,1,1;w)
   &= \frac52 \left[ \frac{4+2\spac e^{-w}}{w^2} 
    - \frac{3\sqrt{\pi}\,\text{erf}\spac\big(\sqrt{w}\big)}{w^{5/2}} \right] , \\
   \Sigma(\textstyle{\frac12},1,1,1;w)
   &= \frac{15\sqrt2}{2\spac w^2}\,\bigg[ 2\sqrt{2} - 2\spac\ln(1+\sqrt{2}) 
    - 4i\pi\,T\bigg(\!\sqrt{2w},\frac{i}{\sqrt2}\bigg) \bigg] \\
   &\quad - \frac{30\sqrt{\pi}}{w^{5/2}} \left[ \text{erf}\spac\big(\sqrt{w}\big)
    - \frac{1}{\sqrt2}\,\text{erf}\spac\bigg(\sqrt{\frac{w}{2}}\bigg) \right] , \\
   \Sigma(0,1,1,1;w)
   &= \frac{15}{w^2} - \frac{15\sqrt{\pi}}{2\spac w^{5/2}}
    \left[ 2\,\text{erf}\spac\big(\sqrt{w}\big) - e^{-w}\,\text{erfi}\spac(\sqrt{w}\big) \right] , \\
   \Sigma(1,1,\textstyle{\frac12},1;w)
   &= \frac{30\sqrt{2}}{w^2} \left[ 2\spac\ln(1+\sqrt{2}) - \sqrt{2} 
    + 4i\pi\,T\bigg(\!\sqrt{2w},\frac{i}{\sqrt2}\bigg) \right] \\
   &\quad + \frac{30\sqrt{2\pi}}{w^{5/2}} \left[ \frac{1}{\sqrt2}\,\text{erf}\spac\big(\sqrt{w}\big)
    - \text{erf}\spac\bigg(\sqrt{\frac{w}{2}}\bigg) 
    + e^{-w}\,\text{erfi}\spac\bigg(\sqrt{\frac{w}{2}}\bigg) \right] .
\end{aligned}
\end{equation}
For the remaining two functions we have obtained one-dimensional integral representations, which can readily be evaluated numerically. They are 
\begin{equation}
\begin{aligned}
   \Sigma(\textstyle{\frac12},1,\textstyle{\frac12},1;w)
   &= \int_0^1\!dz\,\Bigg[ \frac{30\sqrt{\pi}}{\sqrt{w}}\,e^{\frac12\spac w\spac z^2} 
    \left[ \text{erf}\spac\big(\sqrt{w}\big) - \text{erf}\spac\big(\sqrt{w}\spac z\big) \right]
    - \frac{60}{w}\,e^{-\frac{w}{2}}\spac\sinh\!\Big(\frac{w\spac(1-z^2)}{2}\Big) \Bigg] \\
   &\quad\times z^2\,\spac{}_2F_2(1,1; \textstyle{\frac32},2; - \frac12\spac w\spac z^2) \,, \\
   \Sigma(0,1,\textstyle{\frac12},1;w)
   &= \int_0^1\!dz\,\Bigg[ \frac{30\sqrt{\pi}}{w^{3/2}}\,e^{\frac12\spac w\spac z^2} 
    \left[ \text{erf}\spac\big(\sqrt{w}\big) - \text{erf}\spac\big(\sqrt{w}\spac z\big) \right]
    - \frac{60}{w^2}\,e^{-\frac{w}{2}}\spac\sinh\!\Big(\frac{w(1-z^2)}{2}\Big) \Bigg] \\
   &\quad\times \sqrt2 \left[ \ln(1+\sqrt2) 
    + 2i\pi \left( T\big(\sqrt{w}\spac z,i\sqrt2-\varepsilon\big) + \frac14 \right) \right] .
\end{aligned}
\end{equation}

To derive the asymptotic forms of these functions for $w\gg 1$, we start from the expression
\begin{equation}\label{eq:asympt}
\begin{aligned}
   \Sigma(v^{(1)},v^{(2)},v^{(3)},v^{(4)};w)
   &= 5! \int_0^1\!dz_4 \int_0^{z_4}\!dz_3 \int_0^{z_3}\!dz_2 \int_0^{z_2}\!dz_1\,(1-z_4) \\
   &\quad\times e^{-v^{(1)} w\spac z_1^2}\,e^{-v^{(2)} w\spac(z_2^2-z_1^2)}\,
    e^{-v^{(3)} w\spac(z_3^2-z_2^2)}\,e^{-v^{(4)} w\spac(z_4^2-z_3^2)} \,,
\end{aligned}
\end{equation}
which is obtained from~\eqref{eq:Sigmadef} by rescaling $L_i\equiv z_i\spac L_s$. For $w\gg 1$, these integrals can be calculated using the method of regions~\cite{Beneke:1997zp}. In general, they receive contributions from the ``soft'' region $z_4\sim w^{-1/2}\ll 1$ and the ``hard'' region $z_4\sim 1$. However, we find that in the cases where all $v^{(i)}$ values are non-zero the hard region gives rise to exponentially suppressed contributions, while for $v^{(1)}=0$ it contributes terms starting at $\mathcal{O}(1/w^3)$, which are suppressed relative to the two leading terms scaling as $1/w^2$ and $1/w^{5/2}$, respectively.\footnote{For $l=2$ the hard region gives unsuppressed contributions for $v^{(1)}=0$. This fact is responsible for the $\ln(4w)/w$ term and the absence of a term proportional to $w^{-3/2}$ in the expression for $\Sigma(0,1;w)$ in~\eqref{eq:l2asy}.}  
It thus suffices to focus on the soft region, for which the upper limit on the integral over $z_4$ must be replaced by infinity. Introducing new integration variables via the substitutions $z_1=t_1\spac z_2$, $z_2=t_2\spac z_3$ and $z_3=t_3\spac z_4$, and performing the integral over $z_4$, we then find up to higher-order terms
\begin{equation}
   \Sigma(v^{(1)},v^{(2)},v^{(3)},v^{(4)};w) \big|_{w\gg 1}
   = 60 \int_0^1\!dt_1 \int_0^1\!dt_2\spac t_2 \int_0^1\!dt_3\spac t_3^2
    \left[ \frac{1}{w^2}\,\frac{1}{\Delta^2} - \frac{3\sqrt{\pi}}{4\spac w^{5/2}}\,\frac{1}{\Delta^{5/2}} \right] 
    + \dots \,,
\end{equation}
with
\begin{equation}
   \Delta = v^{(1)} t_1^2\,t_2^2\,t_3^2 + v^{(2)} (1-t_1^2)\,t_2^2\,t_3^3 + v^{(3)} (1-t_2^2)\,t_3^2
    + v^{(4)} (1-t_3^2) \,.
\end{equation}
The integral over $t_2$ is straightforward to evaluate. Performing the remaining parameter integrals for the cases of interest, we find the asymptotic forms
\begin{equation}
\begin{aligned}
   \Sigma(1,1,1,1;w)
   &= \frac{10}{w^2} - \frac{15\sqrt{\pi}}{2\spac w^{5/2}} + \mathcal{O}(e^{-w}) \,, \\
   \Sigma(\textstyle{\frac12},1,1,1;w)
   &= \frac{15 \left[ 2 - \sqrt2\spac\ln(1+\sqrt2) \right]}{w^2} 
    - \frac{15\sqrt{\pi}\,(2-\sqrt2)}{w^{5/2}} + \mathcal{O}(e^{-\frac{w}{2}}) \,, \\
   \Sigma(0,1,1,1;w)
   &= \frac{15}{w^2} - \frac{15\sqrt{\pi}}{w^{5/2}} + \mathcal{O}(w^{-3}) \,, \\
   \Sigma(1,1,\textstyle{\frac12},1;w)
   &= \frac{60}{w^2} \left[ \sqrt2\spac\ln(1+\sqrt2) - 1 \right]
    - \frac{30\sqrt{\pi}\,(\sqrt2-1)}{w^{5/2}} + \mathcal{O}(e^{-\frac{w}{2}}) \,, \\
   \Sigma(\textstyle{\frac12},1,\textstyle{\frac12},1;w)
   &= \frac{15\sqrt2}{w^2} \left[ \frac{5\pi^2}{4} - \frac32 \ln^2 2
    - 12\,\text{Li}_2\bigg(\frac{1}{\sqrt2}\bigg) \right] 
    - \frac{15\sqrt{2\pi}\,\ln 2}{w^{5/2}} + \mathcal{O}(e^{-\frac{w}{2}}) \,, \\
   \Sigma(0,1,\textstyle{\frac12},1;w)
   &= \frac{30\spac\ln^2(1+\sqrt2)}{w^2} - \frac{30\sqrt{\pi}\,\ln(1+\sqrt2)}{w^{5/2}} 
    + \mathcal{O}(w^{-3}) \,.
\end{aligned}
\end{equation}

\begin{figure}[t]
\centering
\includegraphics[scale=1]{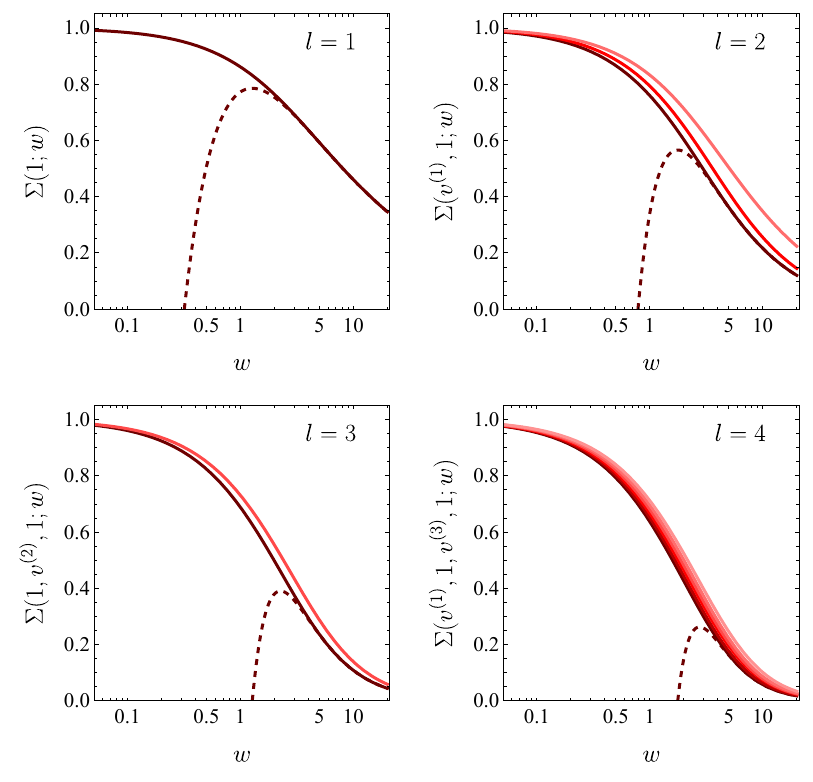}
\caption{The functions $\Sigma(v^{(1)},\dots,v^{(l)}; w)$ for $l\le 4$ are shown from bottom to top in each panel in the order in which they are presented in the text. The relevant values are $v^{(1)}\in\{1,\frac12,0\}$ for $l=2$, $v^{(2)}\in\{1,\frac12\}$ for $l=3$, and $(v^{(1)},v^{(3)})\in\{(1,1),(\frac12,1),(0,1),(1,\frac12),(\frac12,\frac12),(0,\frac12)\}$ for $l=4$. The dashed lines show the asymptotic large-$w$ behavior for the case of $\Sigma(1,\dots,1; w)$.}
\label{fig:Sigma_functions}
\end{figure}

In Figure~\ref{fig:Sigma_functions} we show the relevant functions $\Sigma(v^{(1)},\dots,v^{(l)}; w)$ for $1\le l\le 4$. As a representative example, the dashed line shows the asymptotic behavior for large $w$ for the case of the function $\Sigma(1,\dots,1; w)$. We observe that for higher values of $l$, the asymptotic forms start providing a good approximation at increasingly larger $w$ values. Note also that for given $l$ the differences of two functions belonging to different $v^{(i)}$ values are much smaller than the individual functions. As a consequence, we find that the coefficients of the color operators $\bm{X}_4$ and $\bm{X}_5$ are considerably smaller in magnitude than those of the operators $\bm{X}_2$ and $\bm{X}_3$, cf.~\eqref{eq:USLL2_fixed} and~\eqref{eq:USLL4_fixed}.

\subsubsection*{All-order asymptotic behavior}

The technique described for the case $l=4$ can be straightforwardly extended to higher values of $l$. Focusing on the functions needed in~\eqref{eq:USLL_fixed_coupling_def}, it follows that the leading asymptotic behavior for $w\gg 1$ is
\begin{equation}
   \Sigma(v^{(1)},\dots,v^{(l)}; w)\sim \frac{1}{w^{\spac l/2}} \,,
\end{equation}
with the single exception that for $\Sigma(0,1; w)$ there is an extra factor $\ln w$ in the numerator, which enters only in the fifth component of the evolution vector. For the generic case, we find
\begin{equation}
   \mathbbm{U}_{\rm SLL}^{(l)}(\mu_h,\mu_s) 
   \sim \frac{\left( i\pi \right)^l}{(l+1)!}\,\spac N_c^{l-1} \left( \frac{\alpha_s\spac L_s}{\pi} \right)^{l+1}
    \frac{1}{w^{\spac l/2}}
   = \frac{i^{\spac l}}{(l+1)!}\,\frac{\alpha_s\spac L_s}{\pi\spac N_c}\,\spac w_\pi^{\spac l/2} \,.
\end{equation}
Note the remarkable fact that the dependence on $w$ cancels in the asymptotic limit. In the conventional counting scheme, where $\alpha_s L_s=\mathcal{O}(1)$, higher-order Glauber exchanges are parametrically suppressed in addition to the factorial suppression, since $w\sim1/\alpha_s$ and $w_\pi\sim\alpha_s$. In the alternative counting scheme $\alpha_s L_s^2=\mathcal{O}(1)$~\cite{Becher:2023mtx}, the suppression is even stronger. This reinforces the observation made in~\cite{Boer:2023jsy,Boer:2023ljq} regarding the rapid convergence of the Glauber series, and justifies the suggested truncation after $l=4$ insertions.

\section{Numerical results for quark-initiated scattering}
\label{sec:numerics}

\begin{figure}[t]
\centering
\includegraphics[scale=1]{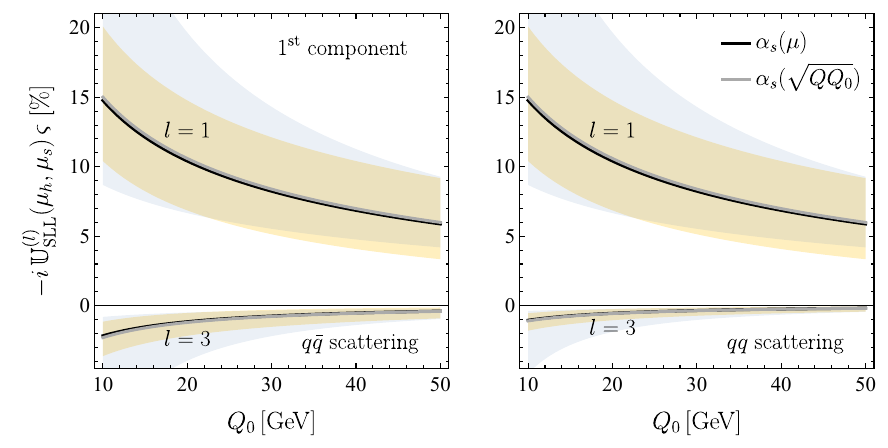}
\vspace{-1mm}
\caption{First component of the evolution vector $\mathbbm{U}_{\rm SLL}^{(l)}(\mu_h,\mu_s)\,\varsigma$ for $q\bar q$ (left) and $qq$ (right) scattering, obtained with $\mu_h=Q$ and $\mu_s=Q_0$, using RG-improved perturbation theory (black lines) or a fixed coupling $\alpha_s(\bar\mu)$ with $\bar\mu=\sqrt{Q\spac Q_0}$ (gray lines). We fix $Q=1$\,TeV and vary the veto scale $Q_0$ as indicated. The yellow bands are obtained from the variation of the low scale $\mu_s$ by a factor of~2 about its default value. The light-blue band shows the estimate of the scale uncertainty from previous works~\cite{Becher:2023mtx,Boer:2023jsy,Boer:2023ljq}, obtained by varying the scale in $\alpha_s(\bar\mu)$ in the interval $\bar\mu\in[Q_0,Q]$.}
\label{fig:UL1}
\end{figure}

\begin{figure}[t]
\centering
\includegraphics[scale=1]{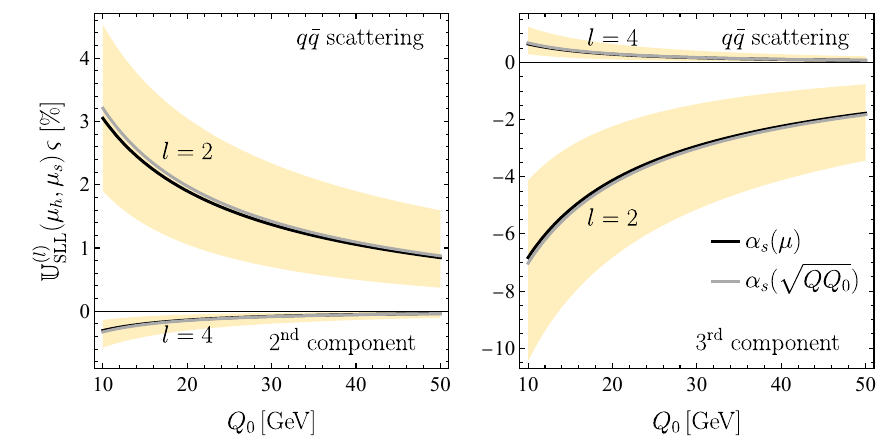}
\vspace{-1mm}
\includegraphics[scale=1]{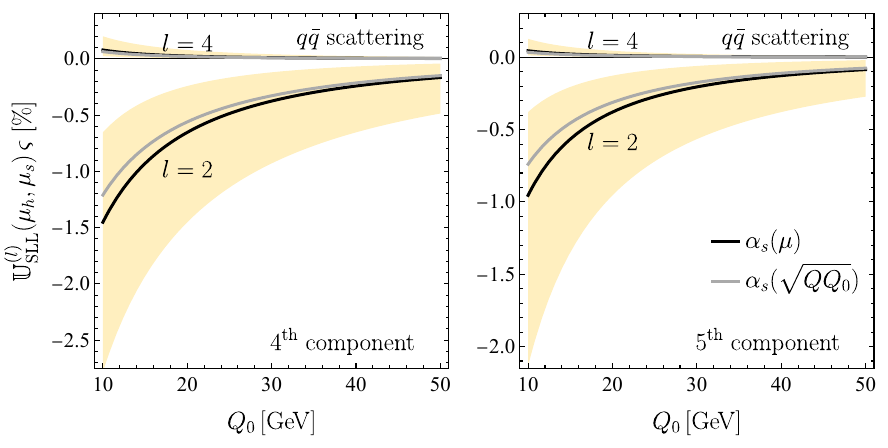}
\vspace{-1mm}
\caption{Components 2\spac--\spac 5 of the evolution vector $\mathbbm{U}_{\rm SLL}^{(l)}(\mu_h,\mu_s)\,\varsigma$ for $q\bar q$ scattering. The color coding and input parameters are the same as in Figure~\ref{fig:UL1}.}
\label{fig:UL2and3}
\end{figure}

Relation~\eqref{eq:sigmaSLLG} shows that the contributions of the Glauber series to a given scattering process is encoded in traces of the five color structures $\bm{X}_i$ with the relevant hard functions and the components of the evolution vectors $\mathbbm{U}_{\rm SLL}^{(l)}(\mu_h,\mu_s)\,\varsigma$ for different values of $l$. Whereas the traces $\langle\bm{\mathcal{H}}_{2\to M}(\mu_h)\,\bm{X}_i\rangle$ are process dependent, the coefficients of the evolution vector, which contain the resummation of the large logarithms, are the same for all $q\bar q$-initiated and all $qq$- or $\bar q\bar q$-initiated scattering processes, irrespective of the flavor of the quarks.

In Figures~\ref{fig:UL1} and \ref{fig:UL2and3} we present numerical results for these coefficients up to $l=4$ derived in RG-improved perturbation theory (black lines). In addition, we show the corresponding curves obtained using a fixed coupling $\alpha_s(\bar\mu)$ with $\bar\mu=\sqrt{\mu_h\spac\mu_s}$ (gray lines), including the two-loop cusp anomalous dimension as done in~\cite{Becher:2023mtx,Boer:2023jsy,Boer:2023ljq}. We choose a fixed value $\mu_h=Q=1$\,TeV for the high-energy scale, which is of order of a typical partonic center-of-mass energy at the LHC, and vary the low scale in the interval $Q_0/2<\mu_s<2\spac Q_0$. The scale variation provides a standard estimate of the perturbative uncertainty and is shown as a yellow band in the plots. Additionally, in Figure~\ref{fig:UL1} the uncertainty estimate adopted in previous works~\cite{Becher:2023mtx,Boer:2023jsy,Boer:2023ljq}, which was obtained by varying the strong coupling in the range between $\alpha_s(Q)$ and $\alpha_s(Q_0)$, is depicted by the light-blue band. This representative example demonstrates that the systematic treatment of the scale uncertainty in RG-improved perturbation theory marks an important improvement over the previous rather conservative estimates. 

Since the higher-order terms in the Glauber series (with $l\ge 3$) receive contributions proportional to $\delta_{q\bar q}$, we differentiate the cases of $q\bar q$ and $qq$ scattering. In general, the higher-order Glauber contributions for $qq$ scattering are smaller than those for $q\bar q$ scattering by a factor of approximately $4/N_c^2$, and we only show them for the first component in the right panel of Figure~\ref{fig:UL1}. In all cases the black and gray lines lie close to each other, indicating that the RG-improved results can be well approximated by using a fixed coupling $\alpha_s(\bar\mu)$ evaluated at the geometric mean $\bar\mu=\sqrt{Q\spac Q_0}$. This fact was already observed for the SLLs in~\cite{Becher:2023mtx}. Next, we observe that the size of the corrections is significantly decreased as one increases the value of $l$. The contributions of higher-order Glauber terms ($l=3$ vs.\ $l=1$, and $l=4$ vs.\ $l=2$) are typically suppressed by an order of magnitude and always have opposite sign, in complete agreement with the findings of~\cite{Boer:2023jsy}. This is in accordance with our finding that in the large-$N_c$ limit the higher Glauber contributions arise only for $q\bar q$ scattering~\cite{Boer:2023jsy}.

As a final comment, we note that the components~4 and 5 of the evolution vector take significantly smaller values than the components~2 and 3. This can be understood from the fact that the expressions for the two lowest components involve differences of two $U_c$ or $\Sigma$ functions belonging to different $v^{(i)}$ values, see~\eqref{eq:USLL2} and~\eqref{eq:USLL4} or~\eqref{eq:USLL2_fixed} and~\eqref{eq:USLL4_fixed}, respectively. These differences are suppressed, because the different functions exhibit a rather similar behavior, as shown in Figure~\ref{fig:Sigma_functions}.

\section{Resummation of SLLs for arbitrary processes}
\label{sec:resummation_arbitrary}

While most of this work has focused on partonic scattering processes with quarks and/or anti-quarks in the initial state, the formalism can be extended to processes involving gluons, see Appendix~\ref{app:gluons}. In this case, the resummation of higher-order terms in the Glauber series requires larger operator bases, containing 14 operators for quark-gluon-initiated scattering processes and 20 structures for gluon-initiated processes. However, if one is only interested in the dominant super-leading logarithms, those arising for $l\le 2$, then a process-independent set consisting of 11 operators is sufficient. They read
\begin{equation} \label{eq:Xi_defs}
\begin{aligned}
   \bm{X}_1 &= \sum_{j>2} J_j\,if^{abc}\,\bm{T}_1^a\spac\bm{T}_2^b\spac\bm{T}_j^c \,,
   &\qquad \bm{X}_6 &= \frac{1}{N_c^3}\,J_{12}\,\bm{S}_1 \,,
   \\
   \bm{X}_2 &= \frac{1}{N_c} \sum_{j>2} J_j\,\bm{O}^{(j)}_1 \,,
   & \bm{X}_7 &= \frac{1}{N_c^3} \, J_{12} \, \bm{S}_2 \,,
   \\
   \bm{X}_3 &= \frac{1}{N_c} \sum_{j>2} J_j\,\bm{O}^{(j)}_2 \,,
   & \bm{X}_8 &= \frac{1}{N_c^3} \, J_{12} \, \bm{S}_3 \,,
   \\ 
   \bm{X}_4 &= \frac{1}{N_c^2} \sum_{j>2} J_j\,\bm{O}^{(j)}_3 \,,
   & \bm{X}_9 &= \frac{1}{N_c^2} \, J_{12} \, \bm{S}_4 \,,
   \\ 
   \bm{X}_5 &= \frac{1}{N_c^2} \sum_{j>2} J_j\,\bm{O}^{(j)}_4 \,,
   & \bm{X}_{10} &= \frac{1}{N_c} \, J_{12} \, \bm{S}_5 \,,
   \\
   && \bm{X}_{11} &= J_{12} \, \bm{S}_6 \,,
\end{aligned}
\end{equation}
where the definitions of $\bm{O}_i^{(j)}$ and $\bm{S}_i$ can be found in~\cite{Becher:2023mtx}. The powers of $1/N_c$ are chosen such that the matrix elements $\langle\bm{\mathcal{H}}_{2\to M}\spac\bm{X}_i\rangle$ are at most of $\mathcal{O}(1)$ in the large-$N_c$ expansion. This set of color structures is closed under repeated applications of $\bm{\Gamma}^c$ for particles transforming in arbitrary representations of $SU(N_c)$, and from \eqref{eq:matrix_reps} one finds the matrix representation
\begin{equation}\label{eq:GGammac11}
   \GGammac = \begin{pmatrix}
    1~ & 0 & 0 & 0 & 0 & 0 & ~~0~~ & ~~0~~ & ~~0~~ & ~~0~~ & ~~0 \\
    0~ & \frac{3}{2} & 0 & \frac{1}{N_c^2} & 0 & 0 & ~~0~~ & ~~0~~ & ~~0~~ & ~~0~~ & ~~0 \\
    0~ & \frac{1}{4} & 1 & 0 & 0 & 0 & ~~0~~ & ~~0~~ & ~~0~~ & ~~0~~ & ~~0 \\
    0~ & 1 & 0 & \frac{3}{2} & 0 & 0 & ~~0~~ & ~~0~~ & ~~0~~ & ~~0~~ & ~~0 \\
    0~ & 1 & 0 & 0 & \frac{1}{2} & 0 & ~~0~~ & ~~0~~ & ~~0~~ & ~~0~~ & ~~0 \\
    0~ & - \frac{N_c^2}{2} & 0 & - 1 & 0 & 2 & ~~0~~ & ~~0~~ & ~~2~~ & ~~0~~ & ~~0 \\
    0~ & - \frac{N_c^2}{4} & 0 & 0 & 0 & \frac{1}{2} & ~~1~~ & ~~0~~ & ~~0~~ & ~~0~~ & ~~0 \\
    0~ & - \frac{N_c^2}{2} & 0 & 0 & 0 & 0 & ~~0~~ & ~~1~~ & ~~0~~ & ~~0~~ & ~~0 \\
    0~ & - 1 & 0 & - \frac{1}{2} & 0 & \frac{2}{N_c^2} & ~~0~~ & ~~0~~ & ~~2~~ & ~~0~~ & ~~0 \\
    0~ & \frac{N_c^2+8}{6} \!-\! \frac{4(C_1+C_2)}{N_c} & ~0~ & \frac{N_c-C_1-C_2}{N_c} & \frac{C_1+C_2}{N_c}
     & 0 & ~~0~~ & ~~0~~ & ~~0~~ & ~~1~~ & ~~0 \\
    0~ & - \frac{4\spac C_1 C_2}{N_c^2} & 0 & 0 & \frac{2\spac C_1 C_2}{N_c^2} & ~\frac{8\spac C_1 C_2}{N_c^4}~
     & ~~0~~ & ~~0~~ & ~~0~~ & ~~0~~ & ~~0
   \end{pmatrix} ,
\end{equation}
where $C_i=C_A=N_c$ if parton $i$ is a gluon and $C_i=C_F=\frac{N_c^2-1}{2N_c}$ if it is a quark or anti-quark. The ``super-leading'' terms in $N_c$, i.e.\ the four entries proportional to $N_c^2$ in the second column, appear worrisome at first sight. However, we show below that they cancel out in all physical results due to two relations between the color structures $\bm{X}_i$ with $i=6,7,8,10$, which are valid in the large-$N_c$ limit. The distinct eigenvalues of $\GGammac$ are \cite{Becher:2023mtx} 
\begin{equation}\label{eq:eigenvalues}
    v_0 = 0 \,, \qquad v_1 = \frac12 \,, \qquad v_2 = 1 \,, \qquad
    v_{3,4} = \frac{3 N_c\pm 2}{2 N_c} \,, \qquad v_{5,6} = \frac{2\spac(N_c\pm 1)}{N_c} \,,
\end{equation}
where $v_3$ and $v_5$ correspond to the plus signs. They appear as the arguments of the functions $U_c(v_i,1; \mu_h,\mu_1,\mu_2)$ in the expression for the evolution operator $\mathbbm{U}_{\rm SLL}^{(2)}(\mu_h,\mu_s)$ given in \eqref{eq:l2solu} below.

We stress that the set $\{\bm{X}_i\}$ is \emph{not} closed under repeated applications of the Glauber operator $\bm{V}^G$, as this would require an extension of the basis of color structures \cite{Boer:2023jsy,Boer:2023ljq}. It is thus not meaningful to write down a matrix representation $\mathbbm{V}^G$ for the Glauber operator on the set $\{\bm{X}_i\}$. However, as we are interested in the SLLs in this section, only two $\bm{V}^G$ insertions are required. The first insertion gives rise to the structure $\bm{X}_1$, see \eqref{eq:firstGlauber}, as encoded in the vector $\varsigma$ in \eqref{eq:sigmaSLLG}. Using the fact that $\varsigma$ is an eigenvector of $\mathbbm{U}_c$, we thus only need to know how the second Glauber operator acts on $\bm{X}_1$. The relevant relation is \cite{Becher:2023mtx}
\begin{equation}
    \bm{V}^G\!: \quad \bm{X}_1 \mapsto i\pi\spac N_c\,\bm{X}_2 \,.
\end{equation}

It is now straightforward to generalize the relations \eqref{eq:USLL1} and \eqref{eq:USLL2} to scattering processes of particles transforming under arbitrary representations of $SU(N_c)$. We find that the result \eqref{eq:USLL1} obtained for a single Glauber-operator insertion remains valid in the general case, while for $l=2$ insertions one obtains
\begin{equation}\label{eq:l2solu}
\begin{aligned}
   \mathbbm{U}_{\rm SLL}^{(2)}(\mu_h,\mu_s)\,\varsigma 
   &= - \frac{32\pi^2}{\beta_0^3}\,N_c \int_1^{x_s}\!\frac{dx_2}{x_2}\,\ln\frac{x_s}{x_2}
    \int_1^{x_2}\!\frac{dx_1}{x_1} \\
   &\hspace{-0.9cm}\times \begin{pmatrix}
	0 \\[2mm]
    \frac12 \left[ U_c(v_3,1) + U_c(v_4,1) \right] \\[4mm]
    - \frac{N_c^2}{2(N_c^2-4)} \left[ U_c(1,1) - \frac{N_c-2}{2N_c}\,U_c(v_3,1)
     - \frac{N_c+2}{2N_c}\,U_c(v_4,1) \right] \\[4mm]
    \frac{N_c}{2} \left[ U_c(v_3,1) - U_c(v_4,1) \right] \\[4mm]
    - \frac{N_c^2}{N_c^2-1} \left[ U_c(\frac12,1) - \frac{N_c-1}{2N_c}\,U_c(v_3,1)
     - \frac{N_c+1}{2N_c}\,U_c(v_4,1) \right] \\[5mm]
    \frac{N_c^2}{2} \left[ U_c(v_3,1) + U_c(v_4,1) - U_c(v_5,1) - U_c(v_6,1) \right] \\[4mm]
    \frac{N_c^4}{4(N_c^2-4)} \left[ \frac{N_c-2}{N_c}\,\big( U_c(v_3,1) - U_c(v_5,1) \big)
     + \frac{N_c+2}{N_c}\,\big( U_c(v_4,1) - U_c(v_6,1) \big) \right] \\[5mm]
    \frac{N_c^4}{N_c^2-4} \left[ U_c(1,1) - \frac{N_c-2}{2N_c}\,U_c(v_3,1)
     - \frac{N_c+2}{2N_c}\,U_c(v_4,1) \right] \\[5mm]
    \frac{N_c}{2} \left[ U_c(v_3,1) - U_c(v_4,1) - U_c(v_5,1) + U_c(v_6,1) \right] \\[5mm]
    \frac{2N_c\spac(C_1+C_2)}{N_c^2-1} \Big[ U_c(\frac12,1) - \frac{(N_c-1)(N_c+2)}{2N_c}\,U_c(v_3,1)
     + \frac{(N_c+1)(N_c-2)}{2N_c}\,U_c(v_4,1) \Big] \\
    - \frac{N_c^2}{3} \Big[ U_c(1,1) - \frac{N_c+4}{2N_c}\,U_c(v_3,1) 
     - \frac{N_c-4}{2N_c}\,U_c(v_4,1) \Big] \\[5mm]
    \frac{4\spac C_1 C_2}{N_c^2-1}\,\Big[ U_c(0,1) - U_c(\frac12,1)
     + \frac{N_c-1}{2N_c}\,\big( U_c(v_3,1) - U_c(v_5,1) \big) \\
    + \frac{N_c+1}{2N_c}\,\big( U_c(v_4,1) - U_c(v_6,1) \big) \Big] 
   \end{pmatrix} ,
\end{aligned}
\end{equation}
where for brevity we have dropped the three scale arguments of $U_c(v_i,1; \mu_h,\mu_1,\mu_2)$. Using the fact that the eigenvalues $v_3$ and $v_4$, as well as $v_5$ and $v_6$, coincide up to subleading terms in the large-$N_c$ limit, one finds that the components 6,7,8,10 in the 11-component vector in the above expression inherit the ``super-leading'' terms in $N_c$ contained in \eqref{eq:GGammac11}. In order for these terms to cancel out in all predictions for physical quantities, the color operators $\bm{X}_6$ and $\bm{X}_7$, as well as $\bm{X}_8$ and $\bm{X}_{10}$, must satisfy the relations
\begin{equation}\label{eq:two_relations}
   \bm{X}_6 + \frac12\,\bm{X}_7 = \mathcal{O}\bigg(\frac{1}{N_c^2}\bigg) \,, \qquad
   \bm{X}_8 - \frac13\,\bm{X}_{10} = \mathcal{O}\bigg(\frac{1}{N_c^2}\bigg) \,.
\end{equation}
The second relation was already established in \cite{Becher:2023mtx} for the case of quarks and gluons, where the structure $\bm{X}_8$ was shown to be redundant. Moreover, for quark-initiated processes it was found that
\begin{equation}
   \bm{X}_6 
   = - \frac{N_c^2}{2(N_c^2-4)}\,\bm{X}_7 - \frac{N_c^2-1}{N_c^4}\,\bm{X}_{11} \,,
\end{equation}
in accordance with the first relation. If at least one of the two initial-state partons is a gluon, the structure $\bm{X}_6$ is not redundant. Assuming that particle~1 is a gluon, and using relations for $SU(N_c)$ group generators compiled in \cite{Haber:2019sgz}, we find that
\begin{equation}\label{eq:X6rela}
   \bm{X}_6 
   = - \frac12\,\bm{X}_7 - \frac{4\spac C_2}{N_c^3}\,\bm{X}_{11} - \frac{2}{N_c^2}\,\bm{X}_6' \,,
\end{equation}
with 
\begin{equation}
   \left[ \bm{X}_6' \right]_{a_1 b_1,a_2\spac b_2}
   = \frac{1}{N_c}\,J_{12}\,\{ \bm{T}_2^{a_1},\bm{T}_2^{b_1} \}_{a_2\spac b_2} \,,
\end{equation}
where the indices $a_i,b_i$ refer to particle~$i$. This establishes the first relation in \eqref{eq:two_relations} for arbitrary initial-state partons in QCD. If one would eliminate the redundant operator $\bm{X}_8$ from the basis and express $\bm{X}_6$ in favor of $\bm{X}_6'$ using \eqref{eq:X6rela}, the ``super-leading'' terms in $N_c$ would be eliminated from the expressions analogous to \eqref{eq:GGammac11} and \eqref{eq:l2solu}. In practice, however, it is more convenient to work with the original definitions of the basis structures.

For completeness, we mention that if the result \eqref{eq:l2solu} is evaluated working with a fixed coupling $\alpha_s(\bar\mu)$, one encounters four new functions $\Sigma(v,1; w)$ with $v>0$, which can be evaluated using~\cite{Becher:2023mtx}
\begin{equation} \label{eq:new_sigma}
\begin{aligned}
   \Sigma(v,1;w) 
   &= \frac{3}{2z\sqrt{w}}\,\bigg[ 4\pi\,T\bigg(\sqrt{2}\spac z,\frac{\sqrt{w}}{z}\bigg)
		- \frac{\sqrt{\pi}\spac z\,\text{erf}\big(\sqrt{v\spac w}\big)}{\sqrt{v}\spac w} 
		+ \frac{\sqrt{\pi}\,e^{-w}\,\text{erf}(z)}{\sqrt{w}} \\ 
   &\hspace{1.9cm} + \pi\,\text{erf}\big(\sqrt{w}\big)\,\text{erf}(z) 
    + 2 \arccos\left(\frac{1}{\sqrt{v}}\right) - \pi \bigg] \\
   &= \frac{3\arctan(\sqrt{v-1})}{\sqrt{v-1}\,w} 
    - \frac{3\sqrt{\pi}}{2\sqrt{v}\,w^{3/2}} + \mathcal{O}(e^{-c\spac w}) \,,
\end{aligned}
\end{equation}
with $z=\sqrt{(v-1+i\varepsilon)\spac w}$ and $c=\min(v,1)$. The asymptotic behavior is most readily derived from eq.\,(5.23) given in~\cite{Becher:2023mtx}.\footnote{In this reference, the neglected higher-order terms were denoted as $\mathcal{O}(w^{-2})$. More specifically, one finds that they are of $\mathcal{O}(w^{-2}\,e^{-c\spac w})$, which we simply write as $\mathcal{O}(e^{-c\spac w})$ in the present work.}

\begin{figure}[t]
\centering
\includegraphics[scale=1]{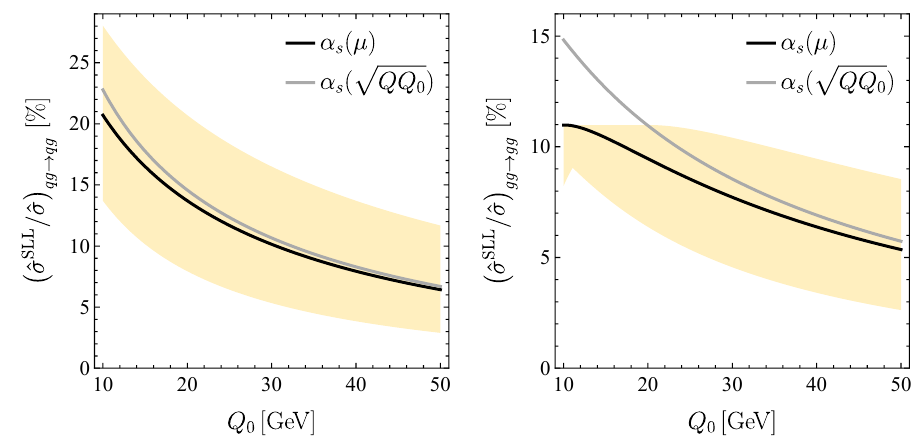}
\vspace{-1mm}
\caption{Numerical results for SLL contributions to partonic $qg\to qg$ forward scattering (left) and $gg\to gg$ small-angle scattering (right) as a function of the jet-veto scale $Q_0$. The meaning of the curves and the yellow bands is the same as in Figure~\ref{fig:UL1}. We choose a rapidity gap $\Delta Y=2$ and $Q=1$\,TeV.}
\label{fig:RG_improved_SLL}
\end{figure}

In Figure~\ref{fig:RG_improved_SLL} we show the resummed contribution of the SLLs, i.e.\ the $l=2$ term of~\eqref{eq:sigmaSLLG}, for two partonic scattering processes with gluons in the initial state, comparing our new results obtained in RG-improved perturbation theory (black lines and yellow bands) with the results derived in \cite{Becher:2023mtx} using a fixed coupling $\alpha_s(\bar\mu)$ with $\bar\mu=\sqrt{Q\spac Q_0}$ (gray lines).
For a central rapidity gap the angular integrals in~\eqref{eq:Jints} evaluate to $J_{3,4}=\mp\Delta Y$ and $J_{12}=\Delta Y$. We choose $\Delta Y=2$.
For quark-gluon elastic scattering in the forward region (left panel), the RG-improved result is well approximated by the fixed-coupling approximation; however, RG improvement allows us to perform a meaningful estimate of the scale uncertainty by varying the low scale $\mu_s$ by a factor~2 about its default value (yellow band). In contrast, for $gg\to gg$ small-angle scattering (right panel) the fixed-coupling approximation becomes worse at small values of $Q_0$, necessitating an RG-improved treatment using a running coupling $\alpha_s(\mu)$ inside the scale integrals. The band indicating the scale uncertainty in the right plot has a somewhat special shape for $Q_0<20$\,GeV, because the SLL contribution peaks near $\mu_s\approx Q_0$ in this region.

\section{Conclusions}
\label{sec:conclusions}

In this paper we have reformulated the resummation of super-leading logarithms and the Glauber series for non-global LHC observables in such a way that it can be systematically performed in RG-improved perturbation theory. Compared to the previous works~\cite{Becher:2021zkk,Becher:2023mtx,Boer:2023jsy,Boer:2023ljq}, including the running of the coupling $\alpha_s(\mu)$ constitutes an important step forward toward reliably estimating the perturbative uncertainties and providing a systematically improvable framework for analyzing such contributions. This step is achieved by means of a new strategy that involves treating the Glauber operator $\bm{V}^G$ perturbatively and expanding the evolution operator in \eqref{eq:Glauber_series} accordingly, as shown in \eqref{eq:USLLndef}. This re-ordering results in the collection of all double-logarithmic corrections in the evolution operator $\bm{U}_c(\mu_i,\mu_j)$ in \eqref{eq:Ucmuimuj}, which takes the form of a matrix-valued Sudakov operator. This operator is no longer path-ordered but instead an ordinary matrix exponential, and its evaluation becomes straightforward with the help of the bases of color operators developed in~\cite{Boer:2023jsy,Boer:2023ljq}. The remaining task is then to obtain the relevant coefficient vectors $\mathbbm{U}_{\rm SLL}^{(l)}(\mu_h,\mu_s)\,\varsigma$ in~\eqref{eq:sigmaSLLG}. We have explicitly solved this problem for the case of quark-initiated scattering processes in Section~\ref{sec:Uoperator}, while the ingredients needed for the extension to arbitrary scattering processes featuring also gluons in the initial state are presented in Appendix~\ref{app:gluons}. When one is interested in the super-leading logarithms only ($l=1,2$), one can determine the coefficient vectors using a process-independent set of basis operators, as demonstrated in Section~\ref{sec:resummation_arbitrary}. The $l^{\rm th}$ term in the series of coefficient vectors, where $l$ denotes the number of Glauber-operator insertions, contains $l$ non-trivial scale integrals. We have evaluated them numerically at leading order in RG-improved perturbation theory, i.e.\ using the two-loop expressions of the cusp anomalous dimension and the QCD $\beta$-function for the logarithmically-enhanced terms. This is a significant improvement over previous analyses, which formally required the evaluation of an infinite number of increasingly more complicated scale integrals.

In the approximation where one works with a fixed coupling $\alpha_s(\bar\mu)$, the coefficient vectors can be expressed in terms of functions $\Sigma(v^{(1)},\dots,v^{(l)};w)$ depending on $w\sim\alpha_s\spac L^2$ with $L=\ln(Q/Q_0)$. For the cases $l\le 4$, and restricting the focus to quark-initiated scattering processes, we have provided explicit expressions for these functions in terms of special functions or simple integral representations. Since the running of the coupling is a single-logarithmic effect, one can use the fixed-coupling approximation to determine the asymptotic behavior of subleading logarithmic corrections for arbitrary $l$ in the limit $w\gg 1$. We have shown that in the conventional counting scheme, where $\alpha_s\spac L\sim 1$, the higher-order Glauber contributions are factorially suppressed and parametrically suppressed by $1/w^{l/2}$.

We have presented numerical results for the coefficient vectors at leading order in RG-improved perturbation theory and quantified the perturbative uncertainties through variation of the low scale $\mu_s$ by a factor~2 about the default value $Q_0$. This procedure leads to significantly smaller estimate of the scale uncertainties than previous analyses using a fixed coupling $\alpha_s(\bar\mu)$, where varying the scale $\bar\mu\in[Q_0,Q]$ served as an (overly) conservative error estimate. In most cases, the central value of our predictions is well approximated by using a fixed coupling $\alpha_s(\sqrt{Q\spac Q_0})$.

The new strategy presented here elucidates the structure of the super-leading logarithmic series, revealing that all double logarithms originate from multiple generalized Sudakov factors. While individually these exhibit exponential suppression, the additional scale integrals result in a very different functional dependence of the resummed expressions, which exhibit only a power-like fall-off. This intuitive picture unveils that the super-leading logarithms are not inherently enhanced. Together with the color bases of~\cite{Boer:2023ljq}, our setup provides a convenient RG-based framework for the study of subleading logarithmic corrections, in particular those from multiple soft emissions. 

\subsubsection*{Acknowledgments}

It is a pleasure to thank Thomas Becher for valuable discussions. This research has received funding from the European Research Council (ERC) under the European Union’s Horizon 2022 Research and Innovation Program (ERC Advanced Grant agreement No.101097780, EFT4jets). Views and opinions expressed are however those of the authors only and do not necessarily reflect those of the European Union or the European Research Council Executive Agency. Neither the European Union nor the granting authority can be held responsible for them. The work reported here was also supported by the Cluster of Excellence \textit{Precision Physics, Fundamental Interactions, and Structure of Matter} (PRISMA$^+$, EXC 2118/1) within the German Excellence Strategy (Project-ID 390831469).

\begin{appendix}

\section{Generalization to gluons in the initial state}
\label{app:gluons}

In this appendix, we extend the formalism described in Section~\ref{sec:Uoperator} to include gluons in the initial state. Most results from the main text can be generalized straightforwardly; however, the color bases are now much more involved and the corresponding expressions are more cumbersome. The starting point are the matrix representations of $\bm{\Gamma}^c$ and $\bm{V}^G$ introduced in \eqref{eq:matrix_reps}. When expressed in a basis of color operators, they can be written in the block form
\begin{equation}\label{eq:matrix_decomposition}
   \GGammac 
    = \begin{pmatrix}
     \tilde{\gamma}^{(j)} & 0 & ~0 \\
     0 & \gamma^{(j)} & ~0 \\
     0 & \lambda & ~\gamma
    \end{pmatrix} , \qquad
   \mathbbm{V}^G 
    = \begin{pmatrix}
     0 & \tilde{\nu}^{(j)} & ~0 \\
     \nu^{(j)} & 0 & ~0 \\
     0 & 0 & ~0
    \end{pmatrix} .
\end{equation}
Given the structure of $\GGammac$, the matrix exponential in \eqref{eq:Ucexp} can be calculated in closed form, and hence it is straightforward to evaluate the evolution functions in \eqref{eq:masterints} and \eqref{eq:USLL_fixed_coupling_def}. For details on the construction of the color bases and the decompositions of the matrices we refer to~\cite{Boer:2023ljq}. In the following, we list the relevant basis structures and matrices for quark-gluon- and gluon-initiated processes.

\subsection{Quark-gluon-initiated processes}

In this case the color basis is
\begin{align} \label{eq:color_basis_qg}
	\bm{X}_1 &= \sum_{j>2} J_j \, \bm{O}^{(j)}_{3f,F} \,,
	&\quad
	\bm{X}_4 &= \frac{1}{N_c} \sum_{j>2} J_j \, \bm{O}^{(j)}_{2,F} \,,
	&\quad
	\bm{X}_{12} &= J_{12} \, \bm{O}_0 \,,
	\nonumber\\
	\bm{X}_2 &= \sum_{j>2} J_j \, \bm{O}^{(j)}_{3f,D} \,,
	&\quad
	\bm{X}_5 &= \frac{1}{N_c} \sum_{j>2} J_j \, \bm{O}^{(j)}_{2,T} \,,
	&\quad
	\bm{X}_{13} &= \frac{1}{N_c} \, J_{12} \, \bm{O}_{2,F} \,,
	\nonumber\\
	\bm{X}_3 &= \frac{1}{N_c} \sum_{j>2} J_j \, \bm{O}^{(j)}_{4,\nabla} \,,
	&\quad
	\bm{X}_6 &= \sum_{j>2} J_j \, \bm{O}^{(j)}_{3d,F} \,,
	&\quad
	\bm{X}_{14} &= \frac{1}{N_c} \, J_{12} \, \bm{O}_{2,D} \,,
	\nonumber\\
	&&\quad
	\bm{X}_7 &= \sum_{j>2} J_j \, \bm{O}^{(j)}_{3d,D} \,,
	\nonumber\\
	&&\quad
	\bm{X}_8 &= \frac{1}{N_c} \sum_{j>2} J_j \, \bm{O}^{(j)}_{4,\Delta} \,,
	\nonumber\\
	&&\quad
	\bm{X}_9 &= \frac{1}{N_c^2} \sum_{j>2} J_j \, \bm{O}^{(j)}_{4,FF} \,,
	\nonumber\\
	&&\quad
	\bm{X}_{10} &= \frac{1}{N_c} \sum_{j>2} J_j \, \bm{O}^{(j)}_{2,D} \,,
	\nonumber\\
	&&\quad
	\bm{X}_{11} &= \frac{1}{N_c^2} \sum_{j>2} J_j \, \bm{O}^{(j)}_{4,FD} \,,    
\end{align}
where one can find the definitions of $\bm{O}_i^{(j)}$ and $\bm{O}_i$ in~\cite{Boer:2023ljq}. The powers of $1/N_c$ are chosen such that matrix elements $\langle\bm{\mathcal{H}}_{2\to M}(\mu_h)\,\bm{X}_i\rangle$ for $i=4,5,7,\dots,14$ are at most of $\mathcal{O}(1)$ in the large-$N_c$ expansion. For the remaining structures $i=1,2,3,6$, the powers of $1/N_c$ are determined by imposing that the corresponding entries of the matrices $\GGammac$ and $\mathbbm{V}^G$ are at most of $\mathcal{O}(1)$. The submatrices of $\GGammac$ for quark-gluon-initiated processes are given by
\begin{equation}
\begin{aligned}
	\tilde{\gamma}^{(j)} &=
	\begin{pmatrix}
		1 & 0 & \frac{1}{N_c^2} \\
		0 & 1 & 0 \\
		0 & 0 & \frac{3}{2}
	\end{pmatrix} ,
	&\qquad
	\gamma^{(j)} &=
	\begin{pmatrix}
		\frac{1}{2} & 0 & 0 & 0 & 0 & 0 & 0 & 0 \\
		0 & \frac{1}{2} & 0 & 0 & 0 & -\frac{2}{N_c^2} & 0 & 0 \\
		0 & 0 & 1 & 0 & 0 & 0 & 0 & -\frac{1}{2 N_c^2} \\
		0 & 0 & 0 & 1 & 0 & -\frac{1}{2 N_c^2} & 0 & 0 \\
		0 & 0 & 0 & 0 & \frac{3}{2} & -\frac{1}{N_c^2} & 0 & 0 \\
		0 & 0 & 0 & 0 & -1 & \frac{3}{2} & 0 & 0 \\
		0 & 0 & 0 & 0 & 0 & 0 & \frac{1}{2} & 0 \\
		0 & 0 & 0 & 0 & 0 & 0 & 0 & \frac{3}{2}
	\end{pmatrix} ,
	\\
	\gamma &=
	\begin{pmatrix}
		0 & 0 & 0 \\
		0 & 1 & 0 \\
		0 & 0 & 1
	\end{pmatrix} ,
	&
	\lambda &=
	\begin{pmatrix}
		\frac{1}{2} & -\frac{C_F}{2 N_c} & 0 & 0 & -\frac{1}{2 N_c^2} & -\frac{1}{2 N_c^2} & 0 & 0 \\
		\frac{1}{2} & -\frac{1}{2} & 0 & 0 & -2 & \frac{1}{2} & 0 & 0 \\
		0 & 0 & 0 & 0 & 0 & 0 & \frac{1}{2} & -\frac{1}{2}
	\end{pmatrix} ,
\end{aligned}
\end{equation}
and for $\mathbbm{V}^G$ they read
\begin{align}
	\nu^{(j)} &=
	\begin{pmatrix}
		-1 & 0 & \frac{2}{N_c^2} \\
		2 & 0 & 0 \\
		-\frac{1}{2} & \frac{1}{2} & \frac{2}{N_c^2} \\
		\frac{1}{2} & -\frac{1}{2} & 0 \\
		1 & 0 & -1 \\
		0 & 0 & -1 \\
		0 & -1 & 0 \\
		0 & 0 & -1
	\end{pmatrix} ,
	&
	\tilde{\nu}^{(j)} &=
	\begin{pmatrix}
		-\frac{2}{N_c^2} & \frac{2}{N_c^2} & -\frac{N_c^2-4}{2 N_c^2} & \frac{N_c^2-8}{2 N_c^2} & \frac{1}{N_c^2} & \frac{1}{2 N_c^2} & 0 & -\frac{1}{2 N_c^2} \\
		0 & 0 & \frac{1}{2} & -\frac{N_c^2-4}{2 N_c^2} & -\frac{1}{N_c^2} & -\frac{1}{2 N_c^2} & -\frac{2}{N_c^2} & \frac{1}{2 N_c^2} \\
		0 & 0 & 0 & -1 & -1 & 0 & 0 & -\frac{1}{N_c^2}
	\end{pmatrix} .
\end{align}
The evolution matrix~\eqref{eq:Ucdiag} for quark-gluon-initiated processes contains linear combinations of the functions $U_c(v; \mu_i,\mu_j)$ with six distinct eigenvalues, namely $v_{0,1,2,3,4}$ given in \eqref{eq:eigenvalues} as well as
\begin{equation}\label{eq:neweigenvalue}
   v_7 = \frac32 \,.
\end{equation}

\subsection{Gluon-initiated processes}

In this case the color basis is
\begin{align} \label{eq:color_basis_gg}
	\bm{X}_1 &= \sum_{j>2} J_j \, \bm{A}^{(j)}_{3f,F,F} \,,
	&\quad
	\bm{X}_8 &= \frac{1}{N_c} \sum_{j>2} J_j \, \bm{A}^{(j)}_{2,F} \,,
	&\quad
	\bm{X}_{15} &= J_{12} \, \bm{S}_0 \,,
	\nonumber\\
	\bm{X}_2 &= \sum_{j>2} J_j \, \bm{A}^{(j)}_{3f,D,D} \,,
	&\quad
	\bm{X}_9 &= \sum_{j>2} J_j \, \bm{A}^{(j)}_{3d,F,D} \,,
	&\quad
	\bm{X}_{16} &= \frac{1}{N_c} \, J_{12} \, \bm{S}_{2,F,F} \,,
	\nonumber\\
	\bm{X}_3 &= \frac{1}{N_c} \sum_{j>2} J_j \, \bm{A}^{(j)}_{4,F,\nabla} \,,
	&\quad
	\bm{X}_{10} &= \frac{1}{N_c} \sum_{j>2} J_j \, \bm{A}^{(j)}_{4,F,\Delta} \,,
	&\quad
	\bm{X}_{17} &= \frac{1}{N_c} \, J_{12} \, \bm{S}_{2,D,D} \,,
	\nonumber\\
	\bm{X}_4 &= \frac{1}{N_c} \sum_{j>2} J_j \, \bm{A}^{(j)}_{5f,\Delta,\Delta} \,,
	&\quad
	\bm{X}_{11} &= \frac{1}{N_c^2} \sum_{j>2} J_j \, \bm{A}^{(j)}_{4,F,FF} \,,
	&\quad
	\bm{X}_{18} &= \frac{1}{N_c^2} \, J_{12} \, \bm{S}_{4,\Delta,\Delta} \,,
	\nonumber\\
	\bm{X}_5 &= \frac{1}{N_c^2} \sum_{j>2} J_j \, \bm{A}^{(j)}_{5f,\Delta,FF} \,,
	&\quad
	\bm{X}_{12} &= \frac{1}{N_c^2} \sum_{j>2} J_j \, \bm{A}^{(j)}_{4,D,FD} \,,
	&\quad
	\bm{X}_{19} &= \frac{1}{N_c} \, J_{12} \, \bm{S}_{4,\Delta,FF} \,,
	\nonumber\\
	\bm{X}_6 &= \frac{1}{N_c} \sum_{j>2} J_j \, \bm{A}^{(j)}_{5f,\nabla,\nabla} \,,
	&\quad
	\bm{X}_{13} &= \frac{1}{N_c} \sum_{j>2} J_j \, \bm{A}^{(j)}_{5f,\Delta,\nabla} \,,
	&\quad
	\bm{X}_{20} &= \frac{1}{N_c^2} \, J_{12} \, \bm{S}_{4,\nabla,\nabla} \,,
	\nonumber\\
	\bm{X}_7 &= \frac{1}{N_c^2} \sum_{j>2} J_j \, \bm{A}^{(j)}_{5d,\nabla,FD} \,,
	&\quad
	\bm{X}_{14} &= \frac{1}{N_c^2} \sum_{j>2} J_j \, \bm{A}^{(j)}_{5f,\nabla,FF} \,,
\end{align}
where one can find the definitions of $\bm{A}_i^{(j)}$ and $\bm{S}_i$ in~\cite{Boer:2023ljq}. The powers of $1/N_c$ are chosen such that matrix elements $\langle\bm{\mathcal{H}}_{2\to M}(\mu_h)\,\bm{X}_i\rangle$ for $i=8,\dots,20$ are at most of $\mathcal{O}(1)$ in the large-$N_c$ expansion. For the remaining structures $i=1,\dots,7$, the powers of $1/N_c$ are determined by imposing that the corresponding entries of the matrices $\GGammac$ and $\mathbbm{V}^G$ are at most of $\mathcal{O}(1)$. In this basis, the submatrices of $\GGammac$ in~\eqref{eq:matrix_decomposition} are given by
\begin{align}
	\tilde{\gamma}^{(j)} &=
	\begin{pmatrix}
		1 & 0 & \frac{2}{N_c^2} & 0 & 0 & \frac{2}{N_c^2} & \frac{2}{N_c^2} \\
		0 & 1 & 0 & 0 & -\frac{2}{N_c^2} & 0 & 0 \\
		0 & 0 & \frac{3}{2} & 0 & 0 & 0 & \frac{2}{N_c^2} \\
		0 & 0 & 0 & 2 & -\frac{4}{N_c^2} & 0 & 0 \\
		0 & 0 & 0 & -1 & 2 & 0 & 0 \\
		0 & 0 & 0 & 0 & 0 & 2 & 0 \\
		0 & 0 & 0 & 0 & 0 & 0 & 2
	\end{pmatrix} ,
	&\quad
	\gamma^{(j)} &=
	\begin{pmatrix}
		\frac{1}{2} & 0 & 0 & -\frac{2}{N_c^2} & 0 & 0 & \frac{6}{N_c^2} \\
		0 & 1 & 0 & -\frac{1}{2 N_c^2} & \frac{1}{2 N_c^2} & 0 & \frac{3}{2 N_c^2} \\
		0 & 0 & \frac{3}{2} & -\frac{1}{N_c^2} & 0 & 0 & \frac{1}{N_c^2} \\
		0 & 0 & -1 & \frac{3}{2} & 0 & 1 & 0 \\
		0 & 0 & 0 & 0 & \frac{3}{2} & 0 & \frac{1}{2} \\
		0 & 0 & 0 & 0 & 0 & 2 & -\frac{1}{N_c^2} \\
		0 & 0 & 0 & 0 & 0 & -1 & 2
	\end{pmatrix} ,
	\nonumber\\
	\gamma &=
	\begin{pmatrix}
		0 & 0 & 0 & 0 & -\frac{16}{N_c^2} & 0 \\
		0 & 1 & 0 & 0 & 0 & \frac{4}{N_c^2} \\
		0 & 0 & 1 & 0 & -4 & 0 \\
		0 & 0 & 0 & 2 & -4 & 0 \\
		0 & 0 & 0 & -\frac{1}{N_c^2} & 2 & 0 \\
		0 & 0 & 0 & 0 & 0 & 2
	\end{pmatrix} ,
	&
	\lambda &=
	\begin{pmatrix}
		-1 & 0 & -\frac{4}{N_c^2} & -\frac{2}{N_c^2} & 0 & 0 & \frac{4}{N_c^2} \\
		-1 & 0 & -4 & 1 & \frac{2}{N_c^2} & -2 & \frac{6}{N_c^2} \\
		0 & 0 & -1 & 0 & -1 & 0 & 1 \\
		0 & 0 & -1 & \frac{1}{2} & 0 & 0 & 1 \\
		0 & 0 & \frac{1}{4} & -\frac{1}{2 N_c^2} & 0 & \frac{1}{2} & 0 \\
		0 & 0 & 0 & 0 & \frac{1}{2} & 0 & -\frac{3}{2}
	\end{pmatrix} ,
\end{align}
and for $\mathbbm{V}^G$ they are
\begin{align}
	\nu^{(j)} &=
	\begin{pmatrix}
		2 & 0 & 0 & 0 & \frac{4}{N_c^2} & 0 & -\frac{8}{N_c^2} \\
		\frac{1}{2} & -\frac{1}{2} & 0 & 0 & \frac{1}{N_c^2} & 0 & -\frac{3}{N_c^2} \\
		1 & 0 & -1 & 0 & -\frac{2}{N_c^2} & 0 & 0 \\
		0 & 0 & -1 & -1 & 0 & 1 & 0 \\
		0 & 0 & 0 & 0 & 1 & 0 & -1 \\
		0 & 0 & 0 & \frac{1}{2} & -\frac{2}{N_c^2} & \frac{1}{2} & -\frac{2}{N_c^2} \\
		0 & 0 & 1 & -1 & 1 & 1 & -1
	\end{pmatrix} ,
	&
	\tilde{\nu}^{(j)} &=
	\begin{pmatrix}
		\frac{4}{N_c^2} & \frac{N_c^2-8}{N_c^2} & \frac{2}{N_c^2} & \frac{1}{N_c^2} & -\frac{1}{N_c^2} & 0 & -\frac{2}{N_c^2} \\
		0 & -1 & 0 & -\frac{1}{N_c^2} & \frac{1}{N_c^2} & 0 & 0 \\
		0 & -1 & -1 & 0 & 0 & 0 & \frac{4}{N_c^2} \\
		0 & 0 & 0 & -\frac{2}{N_c^2} & 0 & 1 & 0 \\
		0 & 0 & -1 & 0 & 0 & 0 & \frac{1}{2} \\
		0 & 0 & 0 & 0 & -\frac{2}{N_c^2} & 1 & \frac{2}{N_c^2} \\
		0 & 0 & 0 & 0 & 0 & 0 & -\frac{1}{2}
	\end{pmatrix} .
\end{align}
The evolution matrix~\eqref{eq:Ucdiag} for gluon-initiated processes contains linear combinations of the functions $U_c(v; \mu_i,\mu_j)$ with eleven distinct eigenvalues, namely $v_{0,1,2,3,4,5,6,7}$ given in \eqref{eq:eigenvalues} and \eqref{eq:neweigenvalue} as well as
\begin{equation}
   v_8 = 2 \,, \qquad v_{9,10} = \frac{2 N_c\pm 1}{N_c} \,,
\end{equation}
where $v_9$ corresponds to the plus sign.

We stress that in the operator bases presented in this appendix there are no redundant structures, and the matrices $\GGammac$ and $\mathbbm{V}^G$ do not contain any ``super-leading terms'' in $N_c$, unlike the situation encountered in Section~\ref{sec:resummation_arbitrary}.

\section{Evaluation of the Sudakov exponents}
\label{app:Sudakov}

The perturbative expansion coefficients of the QCD $\beta$-function and the cusp anomalous dimension at one- and two-loop order are given by~\cite{Korchemskaya:1992je}
\begin{equation}
\begin{aligned}
   \beta_0 &= \frac{11}{3}\,C_A - \frac43\,T_F\spac n_f \,, \qquad 
   &\beta_1 &= \frac{34}{3}\,C_A^2 - \frac{20}{3}\,C_A\spac T_F\spac n_f - 4\spac C_F\spac T_F\spac n_f \,, \\
   \gamma_0 &= 4 \,, 
   &\gamma_1
   &= 4 \left[ \left( \frac{67}{9} - \frac{\pi^2}{3} \right) C_A - \frac{20}{9}\,T_F\spac n_f \right] ,
\end{aligned}
\end{equation}
where $C_A=N_c$, $C_F=\frac{N_c^2-1}{2\spac N_c}$ and $T_F=\frac12$ for the gauge group $SU(N_c)$. For $N_c=3$ colors and $n_f=5$ light quark flavors one has $\frac{\beta_1}{\beta_0}=\frac{116}{23}\approx 5.043$ and $\frac{\gamma_1}{\gamma_0}=\frac{151}{9}-\pi^2\approx 6.908$. We have used these values, along with the two-loop expression for the running coupling $\alpha_s^{(5)}(\mu)$, when deriving our numerical results.

In the realistic scenario where the interval $(\mu_s,\mu_h)$ contains the top-quark threshold $\mu_t\sim m_t$, one should in principle not work with a fixed value $n_f=5$ but instead adjust the number of light quark flavors as a function of the scale. For instance, for the scale integrals over the cusp anomalous dimension in~\eqref{eq:Usll} one should replace
\begin{equation}
\begin{aligned}
   \int_{\mu_j}^{\mu_i}\!\frac{d\mu}{\mu}\,\gamma_{\rm cusp}\big(\alpha_s(\mu)\big)
   &\to \int_{\alpha_s(\mu_j)}^{\alpha_s(\mu_i)}\!d\alpha\,
    \frac{\gamma_{\rm cusp}^{(5)}(\alpha)}{\beta^{(5)}(\alpha)} \,; &
   & \mu_j<\mu_i<\mu_t \,, \\
   &\to \int_{\alpha_s(\mu_j)}^{\alpha_s(\mu_t)}\!d\alpha\,
    \frac{\gamma_{\rm cusp}^{(5)}(\alpha)}{\beta^{(5)}(\alpha)}
    + \int_{\alpha_s(\mu_t)}^{\alpha_s(\mu_i)}\!\!d\alpha\,
    \frac{\gamma_{\rm cusp}^{(6)}(\alpha)}{\beta^{(6)}(\alpha)} \,; &~\,
   & \mu_j<\mu_t<\mu_i \,, \\
   &\to \int_{\alpha_s(\mu_j)}^{\alpha_s(\mu_i)}\!\!d\alpha\,
    \frac{\gamma_{\rm cusp}^{(6)}(\alpha)}{\beta^{(6)}(\alpha)} \,; &
   & \mu_t<\mu_j<\mu_i \,,
\end{aligned}
\end{equation}
where the superscripts denote the value of $n_f$ for the various integrals. The running of $\alpha_s(\mu)$ also depends on the value of $n_f$, as described by the relation $d\alpha_s/d\ln\mu=\beta^{(n_f)}(\alpha_s)$. Note that at two-loop order the coupling is continuous at the quark threshold, such that $\alpha_s^{(5)}(\mu_t)=\alpha_s^{(6)}(\mu_t)$. For simplicity, we refrain from adding a superscript on the running coupling itself, i.e.\ we use $\alpha_s(\mu)\equiv\alpha_s^{(n_f)}(\mu)$.

\begin{figure}
\centering
\includegraphics[scale=1]{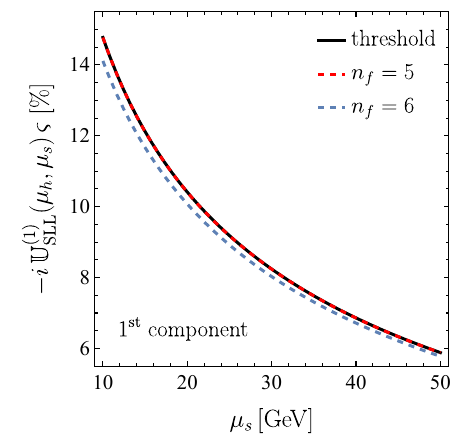}
\caption{Numerical results for the first component of the evolution vector $\mathbbm{U}_{\rm SLL}^{(1)}(\mu_h,\mu_s)\,\varsigma$ using threshold matching (black line), $\alpha_s^{(5)}(\mu)$ (red line) and $\alpha_s^{(6)}(\mu)$ (blue line). The results are obtained with $\mu_h=1$\,TeV and threshold $\mu_t=m_t=175$\,GeV.}
\label{fig:threshold}
\end{figure}

The situation with the generalized Sudakov operator $\bm{U}_c(\mu_i,\mu_j)$ in~\eqref{eq:Ucmuimuj} and the associated scalar function $U_c(v;\mu_i,\mu_j)$ in~\eqref{eq:scalar_evolution_function} is more complicated, because the relevant scale integral
\begin{equation}
   I_h(\mu_i,\mu_j) = \int_{\mu_j}^{\mu_i}\!\frac{d\mu}{\mu}\,\gamma_{\rm cusp}\big(\alpha_s(\mu)\big) \ln\frac{\mu^2}{\mu_h^2} \,, 
\end{equation}
with $\mu_h>\mu_i>\mu_j$, includes a logarithm of the hard scale $\mu_h$. Eliminating the scale $\mu$ in favor of the running coupling $\alpha_s(\mu)$, and distinguishing the different scale hierarchies, we obtain
\begin{align}
    I_h(\mu_i,\mu_j)
    &\to 2 \int_{\alpha_s(\mu_j)}^{\alpha_s(\mu_i)}\!d\alpha\,
     \frac{\gamma_{\rm cusp}^{(5)}(\alpha)}{\beta^{(5)}(\alpha)} 
     \int_{\alpha_s(\mu_h)}^\alpha \frac{d\alpha'}{\beta^{(5)}(\alpha')} \,; &
     & \mu_j<\mu_i<\mu_h<\mu_t \,,
     \nonumber\\ 
    &\to 2 \int_{\alpha_s(\mu_j)}^{\alpha_s(\mu_i)}\!d\alpha\,
    \frac{\gamma_{\rm cusp}^{(5)}(\alpha)}{\beta^{(5)}(\alpha)} 
    \left[ \int_{\alpha_s(\mu_h)}^{\alpha_s(\mu_t)}\!\frac{d\alpha'}{\beta^{(6)}(\alpha')}
    + \int_{\alpha_s(\mu_t)}^\alpha \frac{d\alpha'}{\beta^{(5)}(\alpha')} \right] ; & 
    & \mu_j<\mu_i<\mu_t<\mu_h \,,
    \nonumber\\
    &\to 2 \int_{\alpha_s(\mu_j)}^{\alpha_s(\mu_t)}\!d\alpha\,
    \frac{\gamma_{\rm cusp}^{(5)}(\alpha)}{\beta^{(5)}(\alpha)} 
    \left[ \int_{\alpha_s(\mu_h)}^{\alpha_s(\mu_t)}\!\frac{d\alpha'}{\beta^{(6)}(\alpha')}
    + \int_{\alpha_s(\mu_t)}^\alpha \frac{d\alpha'}{\beta^{(5)}(\alpha')} \right] \nonumber\\*
    &\hspace{5mm} + 2\int_{\alpha_s(\mu_t)}^{\alpha_s(\mu_i)}\!d\alpha\,
    \frac{\gamma_{\rm cusp}^{(6)}(\alpha)}{\beta^{(6)}(\alpha)} 
    \int_{\alpha_s(\mu_h)}^\alpha \frac{d\alpha'}{\beta^{(6)}(\alpha')} \,; &
    & \mu_j<\mu_t<\mu_i<\mu_h \,,
    \nonumber\\
    &\to 2 \int_{\alpha_s(\mu_j)}^{\alpha_s(\mu_i)}\!d\alpha\,
    \frac{\gamma_{\rm cusp}^{(6)}(\alpha)}{\beta^{(6)}(\alpha)} 
    \int_{\alpha_s(\mu_h)}^\alpha \frac{d\alpha'}{\beta^{(6)}(\alpha')} \,; &
    & \mu_t<\mu_j<\mu_i<\mu_h \,.
\end{align}
Expanding the QCD $\beta$-function and the cusp anomalous dimension as shown in~\eqref{eq:pertexp}, it is straightforward to derive from these results the generalization of the expression~\eqref{eq:UcSuda} for the various scale hierarchies. Numerically, it turns out that the effect of accounting for the top-quark threshold is very small. For example, Figure~\ref{fig:threshold} shows $\mathbbm{U}_{\rm SLL}^{(1)}(\mu_h,\mu_s)\,\varsigma$ as a function of $\mu_s$ for the default choices $\mu_h=1$\,TeV and $\mu_t=m_t=175$\,GeV.

\end{appendix}

\clearpage
\pdfbookmark[1]{References}{Refs}
\bibliography{refs.bib}

\end{document}